\documentclass[floatfix, showpacs,showkeys,aps,prd,twocolumn]{revtex4-2}

\usepackage{lineno}
\usepackage{amsmath}
\usepackage{amssymb}

\usepackage{tikz}
\usepackage{lipsum,adjustbox}
\usepackage{graphicx}
\usepackage[caption = false]{subfig}
\usepackage{dcolumn}
\usepackage{bm}
\usepackage{gensymb}

\usepackage[inline]{enumitem}

\let\oldequation\equation
\let\oldendequation\endequation

\renewenvironment{equation}
  {\linenomathNonumbers\oldequation}
  {\oldendequation\endlinenomath}

\newcolumntype{L}[1]{>{\raggedright\arraybackslash}p{#1}}
\newcolumntype{C}[1]{>{\centering\arraybackslash}p{#1}}
\newcolumntype{R}[1]{>{\raggedleft\arraybackslash}p{#1}}

\usepackage{hyperref}
\hypersetup{
 pdftitle={},
 pdfauthor={},
 pdfsubject={},
 pdfkeywords={},
 pdfstartview={},
 bookmarksopen=true, breaklinks=true, debug=true,
 colorlinks=true, linkcolor=blue, citecolor=red, urlcolor=blue,
 hyperfigures=true
}

\newif\ifdraft
\drafttrue

\newcommand{\BESIIIorcid}[1]{\href{https://orcid.org/#1}{\hspace*{0.1em}\raisebox{-0.45ex}{\includegraphics[width=1em]{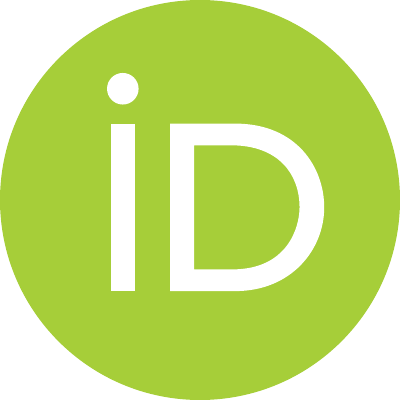}}}}

\begin{document}

\title{\bf \boldmath First measurements of the branching fractions of $J/\psi\to \Xi^0\bar\Lambda K^0_S+c.c.$, $J/\psi\to \Xi^0\bar\Sigma^0 K^0_S+c.c.$, and $J/\psi\to \Xi^0\bar\Sigma^- K^++c.c.$}

\author{
M.~Ablikim$^{1}$\BESIIIorcid{0000-0002-3935-619X},
M.~N.~Achasov$^{4,d}$\BESIIIorcid{0000-0002-9400-8622},
P.~Adlarson$^{81}$\BESIIIorcid{0000-0001-6280-3851},
X.~C.~Ai$^{86}$\BESIIIorcid{0000-0003-3856-2415},
R.~Aliberti$^{38}$\BESIIIorcid{0000-0003-3500-4012},
A.~Amoroso$^{80A,80C}$\BESIIIorcid{0000-0002-3095-8610},
Q.~An$^{63,77,a}$,
Y.~Bai$^{61}$\BESIIIorcid{0000-0001-6593-5665},
O.~Bakina$^{39}$\BESIIIorcid{0009-0005-0719-7461},
Y.~Ban$^{49,i}$\BESIIIorcid{0000-0002-1912-0374}
% ...
\\
\vspace{0.2cm}
(BESIII Collaboration)\\
\vspace{0.2cm} {\it
$^{1}$ Institute of High Energy Physics, Beijing 100049, People's Republic of China\\
% ...
$^{4}$ Budker Institute of Nuclear Physics SB RAS (BINP), Novosibirsk 630090, Russia\\
% ...
$^{38}$ Johannes Gutenberg University of Mainz, Johann-Joachim-Becher-Weg 45, D-55099 Mainz, Germany\\
$^{39}$ Joint Institute for Nuclear Research, 141980 Dubna, Moscow region, Russia\\
% ...
$^{49}$ Peking University, Beijing 100871, People's Republic of China\\
% ...
$^{61}$ Southeast University, Nanjing 211100, People's Republic of China\\
% ...
$^{63}$ State Key Laboratory of Particle Detection and Electronics, Beijing 100049, Hefei 230026, People's Republic of China\\
% ...
$^{77}$ University of Science and Technology of China, Hefei 230026, People's Republic of China\\
% ...
$^{80}$ University of Turin and INFN, (A)University of Turin, I-10125, Turin, Italy; (B)University of Eastern Piedmont, I-15121, Alessandria, Italy; (C)INFN, I-10125, Turin, Italy\\
$^{81}$ Uppsala University, Box 516, SE-75120 Uppsala, Sweden\\
% ...
$^{86}$ Zhengzhou University, Zhengzhou 450001, People's Republic of China\\
% ...
$^{a}$ Deceased\\
% ...
$^{d}$ Also at the Novosibirsk State University, Novosibirsk, 630090, Russia\\
% ...
$^{i}$ Also at State Key Laboratory of Nuclear Physics and Technology, Peking University, Beijing 100871, People's Republic of China\\
% ...
}}

\date{\today}

\begin{abstract}
  By analyzing $(10087 \pm 44)\times10^6$ $J/\psi$ events collected with the BESIII detector at the BEPCII, the decays $J/\psi\to \Xi^0\bar\Lambda K^0_S+c.c.$, $J/\psi\to \Xi^0\bar\Sigma^0 K^0_S+c.c.$, and $J/\psi\to \Xi^0\bar\Sigma^- K^++c.c.$ are observed for the first time. Their branching fractions are determined to be $\mathcal{B}(J/\psi\to \Xi^0\bar\Lambda K^0_S+c.c.)=(3.76\pm0.14\pm 0.22)\times10^{-5}$, $\mathcal{B}(J/\psi\to \Xi^0\bar\Sigma^0 K^0_S+c.c.)=(2.24\pm0.32\pm 0.22)\times10^{-5}$, and $\mathcal{B}(J/\psi\to \Xi^0\bar\Sigma^- K^++c.c.)=(5.64\pm0.17\pm 0.27)\times10^{-5}$, where the first uncertainties are statistical and the second systematic.
\end{abstract}

\maketitle

\section{INTRODUCTION}\label{sec:intro}

Charmonium resonances lie within the transition region 
between the perturbative and non-perturbative regimes of Quantum Chromodynamics (QCD)~\cite{Kwong:1987mj, Eichten:2007qx}.
Below the open charm threshold, both $J/\psi$ and $\psi(3686)$ decay into light hadrons through the annihilation of the $c\bar{c}$ pair into three gluons or one single virtual photon, with the decay width proportional to the square of the charmonium wave function~\cite{aspect3}. While QCD has been extensively tested in the high energy region where the strong interaction coupling constant is small, its perturbative calculations in the low energy region require supplementary non-perturbative contributions. To address this, various effective field theories have been developed~\cite{Eichten:1976jk, Drouffe:1983fv, Chernyak:1983ej}. The study of charmonium decays can provide valuable insights for enhancing our understanding of the internal charmonium structure and for testing phenomenological mechanisms of non-perturbative QCD. Additionally, the violation of the perturbative QCD (pQCD) 12\% rule, namely $\frac{\mathcal{B}(\psi(3686)\to h)}{\mathcal{B}(J/\psi\to h)} = \frac{\mathcal{B}(\psi(3686)\to e^+e^-)}{\mathcal{B}(J/\psi\to e^+e^-)} = 12.7\%$, was first observed by the Mark-II Collaboration in the $\rho\pi$ mode~\cite{aspect4}, which later became known as the $\rho\pi$ puzzle. Since then, numerous decay channels have been employed to investigate the $\rho\pi$ puzzle. Combining the branching fractions of $J/\psi\to \Xi^0\bar\Lambda K^0_S$, $J/\psi\to \Xi^0\bar\Sigma^0 K^0_S$, and $J/\psi\to \Xi^0\bar\Sigma^- K^+$ presented here and future measurements of their counterparts in $\psi(3686)$ decays will provide an additional test of the 12\% rule and improve our understanding of QCD effects in charmonium decays~\cite{bes3-white-paper}.

The decays of $J/\psi \to B\bar{B}^\prime P$, where $B$ denotes baryons and $P$ pseudoscalar mesons, offer a unique platform to search for intermediate states such as baryonia~\cite{Chan:1977qe,Wan:2021vny} and excited baryons that remain unobserved~\cite{Sarantsev:2019xxm}. This is particularly relevant for $\Xi$ hyperons with strangeness $S=-2$. 
More than thirty excited $\Xi$ states have been predicted by theoretical models~\cite{Capstick:2000qj,Glozman:1997ag,Glozman:1995fu,Capstick:1986ter}, but very few of them have been observed and well established due to the small production cross sections and the complicated topology of the final states.

In recent years, partial wave analysis (PWA) has been performed on the $\psi(3686)\to K^-\Lambda \bar{\Xi}^{+}$~\cite{BESIII:2015dvj,BESIII:2023mlv}. However, no studies of $J/\psi[\psi(3686)]\to\Xi^0\bar{B}P$ have been reported to date. In this paper, we present the first observation and branching fraction measurements of the decays $J/\psi\to \Xi^0\bar\Lambda K^0_S$, $J/\psi\to \Xi^0\bar\Sigma^0 K^0_S$, and $J/\psi\to \Xi^0\bar\Sigma^- K^+$, using $(10087 \pm 44)\times10^6$ $J/\psi$ events collected with the BESIII detector~\cite{jpsinumber}. Throughout this paper, charge conjugations are always implicitly included.

\section{BESIII EXPERIMENT AND MONTE CARLO SIMULATION}
The BESIII detector~\cite{BES3} records symmetric $e^+e^-$ collisions provided by the BEPCII storage ring~\cite{Yu:2016cof} in the center-of-mass energy ($\sqrt{s}$) range from 1.84 to 4.95~GeV, with a peak luminosity of $1.1 \times 10^{33}\;\text{cm}^{-2}\text{s}^{-1}$ achieved at $\sqrt{s} = 3.773\;\text{GeV}$.
BESIII has collected large data samples in this energy region~\cite{bes3-white-paper,EcmsMea,EventFilter}. The cylindrical core of the BESIII detector covers 93\% of the full solid angle and consists of a helium-based multilayer drift chamber~(MDC), a time-of-flight system~(TOF), and a CsI(Tl) electromagnetic calorimeter~(EMC), which are all enclosed in a superconducting solenoidal magnet providing a 1.0~T magnetic field. The magnetic field was 0.9~T in 2012, which affects 11\% of the total $J/\psi$ data. The solenoid is supported by an octagonal flux-return yoke with resistive plate counter muon identification modulus interleaved with steel. 
The charged-particle momentum resolution at $1~{\rm GeV}/c$ is $0.5\%$, and the ${\rm d}E/{\rm d}x$ resolution is $6\%$ for electrons from Bhabha scattering. The EMC measures photon energies with a resolution of $2.5\%$ ($5\%$) at $1$~GeV in the barrel (end cap) region. The time resolution in the plastic scintillator TOF barrel region is 68~ps, while that in the end cap region was 110~ps. The end cap TOF system was upgraded in 2015 using multigap resistive plate chamber technology, providing a time resolution of 60~ps, which benefits 87\% of the data used in this analysis~\cite{etof}.

Simulated samples are produced with a {\sc geant4}-based~\cite{geant4} Monte Carlo (MC) package, which includes the geometric description of the BESIII detector and the detector response. The simulation models the beam energy spread and initial state radiation in the $e^+e^-$ annihilations with the generator {\sc kkmc}~\cite{kkmc}. To estimate background contributions, an inclusive MC sample is generated including the production of the $J/\psi$ resonance incorporated in {\sc kkmc}~\cite{kkmc}. All particle decays are modeled with {\sc evtgen}~\cite{evtgen} using branching fractions either taken from the Particle Data Group~(PDG)~\cite{pdg}, when available, or otherwise estimated with {\sc lundcharm}~\cite{lundcharm}. Final state radiation from charged final state particles is incorporated using the {\sc photos} package~\cite{photos}.

The signal decays $J/\psi\to \Xi^0\bar\Lambda K^0_S$ and $J/\psi\to \Xi^0\bar\Sigma^0 K^0_S$ are generated with a phase space (PHSP) model, as no significant intermediate structures are observed. However the PHSP model fails to describe the data in the signal decay $J/\psi\to \Xi^0\bar\Sigma^- K^+$, hence the helicity amplitude model is adopted for simulation~\cite{Chungformalism,Richmanformalism,tfpwajiangyi}, including contributions from the $\Xi^0\bar\Xi^0(1690)$, $\Xi^0\bar\Xi^0(1720)$ and $\Sigma^+(1890)\bar\Sigma^-$. The masses and widths of $\bar\Xi^0(1690)$ and $\Sigma^+(1890)$ are fixed to their respective PDG values~\cite{pdg}, while for $\bar\Xi^0(1720)$, its mass and width are fixed at $m=1.720$~GeV and $\Gamma=0.031$~GeV, respectively.
All subsequent decays of $\Lambda$, $\Sigma^0$, $\Xi^0$, $\Sigma^+$, $K_S^0$, and $\pi^0$ are generated with the PHSP model. These exclusive MC events are used to determine the detection efficiency. Due to the limited statistics and the large systematic uncertainty arising from the low momentum $\Lambda$, the PWA result is not reported and only used to determine the detection efficiency.

In addition, the data sample collected at $\sqrt s=3.080$ GeV with an integrated luminosity of 167.1~pb$^{-1}$~\cite{jpsinumber} is used to estimate the contribution from continuum processes.

\section{Event selection}
The $\Lambda$, $\Sigma^0$, $\Xi^0$, $\Sigma^+$, $K_S^0$, and $\pi^0$ candidates are reconstructed via the $\Lambda \to p\pi^-$, $\Sigma^0 \to \gamma \Lambda$, $\Xi^0\to \Lambda\pi^0$, $\Sigma^+\to p\pi^0$, $K_S^0\to \pi^+\pi^-$, and $\pi^0 \to \gamma \gamma$ decays, respectively.

Charged tracks detected in the MDC are required to be within a polar angle $(\theta)$ range of $|\cos \theta|<0.93$, where $\theta$ is defined with respect to the $z$ axis, which is the symmetry axis of the MDC. For charged tracks not originating from $K_S^0$ or $\Lambda$ decays, the distance of closest approach to the interaction point (IP) must be less than 10\,cm along the $z$ axis, $|V_z|$, and less than 1 cm in the transverse plane, $|V_{xy}|$. Particle identification (PID) for charged tracks combines measurements of the energy deposited in the MDC ($\text{d}E/\text{d}x$) and the flight time in the TOF to form likelihoods $\mathcal{L}(h)(h=p,K,\pi)$ for each hadron $h$ hypothesis. Tracks are identified as protons when the proton hypothesis has the greatest likelihood ($\mathcal{L}(p)>\mathcal{L}(K)$ and $\mathcal{L}(p)>\mathcal{L}(\pi)$), while charged kaons are identified by $\mathcal{L}(K)>\mathcal{L}(\pi)$. No PID is performed on charged pions.

Photon candidates are identified using isolated showers in the EMC. The deposited energy of each shower must be more than 25~MeV in the barrel region ($|\cos \theta|< 0.80$) and more than 50~MeV in the end cap region ($0.86 <|\cos \theta|< 0.92$). To exclude showers that originate from charged tracks, the angle subtended by the EMC shower and the position of the closest charged track at the EMC must be greater than 10 degrees as measured from the IP. To suppress electronic noise and showers unrelated to the event, the difference between the EMC time and the event start time is required to be within [0, 700]\,ns.

Candidates for $K_S^0$ and $\Lambda$ are reconstructed from two opposite charged tracks, each required to satisfy $|V_z| < 20~\text{cm}$. The tracks are assumed to be $\pi^+ \pi^-$ for $K_S^0$, while $p \pi^-$ for $\Lambda$. A common vertex constraint is imposed for both $K_S^0$ and $\Lambda$ candidates, and their decay lengths are required to be greater than twice the vertex resolution away from the IP~\cite{vertex}. The quality of the secondary vertex fit is ensured by a requirement of $\chi^2 < 200$, with no $\chi^2$ condition applied to the primary vertex fit.

The $\pi^0$ candidates are formed by photon pairs with invariant mass in the range $(0.115,\,0.150)$~GeV$/c^{2}$.

A kinematic fit is performed imposing four-momentum conservation (4C) and an additional mass constraint (1C) for each $\pi^0$ in the final state to its nominal mass. This results in a 5C fit for the decay $J/\psi \to (\gamma)\Lambda\pi^0\bar\Lambda K^0_S$ and a 6C fit for $J/\psi \to \Lambda\pi^0\bar{p}\pi^0 K^+$. To improve the agreement in the $\chi^2$ distributions between data and MC simulation, the helix parameters of charged tracks in the MC are corrected using the method described in Ref.~\cite{helixkpi}.
Events are selected based on the following $\chi^2$ requirements: $\chi^2_{\rm{5C}} < 50$ for $J/\psi \to \Lambda\pi^0\bar\Lambda K^0_S$, $\chi^2_{\rm{5C}} < 200$ for $J/\psi \to \gamma\Lambda\pi^0\bar\Lambda K^0_S$, and $\chi^2_{\rm{6C}} < 50$ for $J/\psi \to \Lambda\pi^0\bar{p}\pi^0 K^+$. These thresholds are optimized to maximize the figure of merit, defined as $S / \sqrt{S + B}$, where $S$ and $B$ denote the signal and background yields estimated from the inclusive MC sample, normalized according to the data luminosity and known branching fractions. In events with multiple combinations, only the candidate with the smallest $\chi^2_{\rm{5C}}$ or $\chi^2_{\rm{6C}}$ is retained.

For the decay $J/\psi \to \Lambda\pi^0\bar\Lambda K^0_S$, each candidate event must contain one $\Lambda$ and one $\bar\Lambda$. Since the true $\Lambda$ ($\bar\Lambda$) from $\Xi^0 \to \Lambda\pi^0$ ($\bar\Xi^0 \to \bar\Lambda\pi^0$) must be paired with the corresponding $\pi^0$, two possible assignments exist: one where the $\bar\Lambda$ ($\Lambda$) is prompt, and the other where it is the daughter of a $\Xi^0$ ($\bar\Xi^0$), or vice versa. The correct assignment is determined by minimizing the quantity:
\begin{equation}
\Delta = \sqrt{(M(\Lambda_\text{prompt}) - m_\Lambda)^2 + (M(\pi^0\Lambda_{\Xi^0}) - m_{\Xi^0})^2},
\end{equation}
where $m_\Lambda$ and $m_{\Xi^0}$ are the nominal masses from Ref.~\cite{pdg}. To further reduce background, both $\Lambda$ and $\bar\Lambda$ are required to lie within a mass window of $|M_{p\pi} - m_{\Lambda}| < 7$ MeV/$c^2$, corresponding to a $3\sigma$ region around the nominal $\Lambda$ mass as determined from the signal MC sample.

For the decay $J/\psi \to \gamma\Lambda\pi^0\bar\Lambda K^0_S$, the $\Lambda$ ($\bar\Lambda$) may originate from a $\Sigma^0$ or $\Xi^0$ ($\bar\Sigma^0$ or $\bar\Xi^0$). The assignment is chosen by minimizing:
\begin{equation}
\Delta = \sqrt{(M(\gamma\Lambda_{\Sigma^0}) - m_{\Sigma^0})^2 + (M(\pi^0\Lambda_{\Xi^0}) - m_{\Xi^0})^2},
\end{equation}
with $m_{\Sigma^0}$ taken from Ref.~\cite{pdg}. Background suppression is enhanced by requiring $\Lambda$ candidates to satisfy $|M_{p\pi} - m_{\Lambda}| < 9$ MeV/$c^2$ and $K_S^0$ candidates to satisfy $|M_{\pi^+\pi^-} - m_{K_S^0}| < 7$ MeV/$c^2$, both corresponding to $3\sigma$ intervals based on the signal MC.

In the case of $J/\psi \to \Lambda\pi^0\bar{p}\pi^0 K^+$, the $\bar{\Sigma}^-$ and $\Xi^0$ candidates are selected by minimizing:
\begin{equation}
\Delta = \sqrt{(M(\bar{p}\pi^0_{\bar\Sigma^-}) - m_{\bar\Sigma^-})^2 + (M(\Lambda\pi^0_{\Xi^0}) - m_{\Xi^0})^2},
\end{equation}
where $m_{\bar\Sigma^-}$ is the nominal mass from Ref.~\cite{pdg}. The $\Lambda$ candidate is required to lie within $|M_{p\pi^+} - m_{\Lambda}| < 9$ MeV/$c^2$ ($3\sigma$).

Potential background sources are studied using the inclusive MC sample with TopoAna~\cite{topo}. The dominant background for $J/\psi \to \Xi^0\bar{\Lambda}K_S^0$ is found to be $J/\psi \to \bar{\Xi}^+ \Xi^*(1530)^-$, with $\bar{\Xi}^+ \to \pi^+\bar{\Lambda}$ and $\Xi^*(1530)^- \to \pi^-\Xi^0$. This process peaks in the $\bar{\Lambda}$ and $\Xi^0$ mass distributions but is flat in the $K_S^0$ spectrum. A similar background affects $J/\psi \to \Xi^0\bar{\Sigma}^0K_S^0$, peaking in the $\Xi^0$ mass and flat in $\bar{\Sigma}^0$ and $K_S^0$. For $J/\psi \to \Xi^0\bar{\Sigma}^-K^+$, all backgrounds are combinatorial, with no peaking components. The main combinatorial background is $J/\psi \to \pi^0 K^*(892)^+\bar{p}\Lambda$ (with $K^*(892)^+ \to K^+\pi^0$), accounting for about 50\% of the total. The continuum background is examined using off-resonance data samples; only one event passes the selection for $J/\psi \to \Xi^0\bar{\Lambda}K^0_S$, indicating that continuum contributions are negligible.

\section{Branching fraction measurement}

\subsection{Signal yields in data}

The signal yields of the $J/\psi\to\Xi^0\bar\Lambda K^0_S$, $J/\psi\to\Xi^0\bar\Sigma^0 K^0_S$, and $J/\psi\to\Xi^0\bar\Sigma^- K^+$ decays are determined through an unbinned maximum likelihood fit to their respective two-dimensional (2D) invariant mass distributions: $M_{\pi^+\pi^-}~\text{vs.}~M_{\Lambda\pi^0}$ for $J/\psi\to \Xi^0\bar\Lambda K^0_S$, $M_{\gamma\bar\Lambda}~\text{vs.}~M_{\Lambda\pi^0}$ for $J/\psi\to\Xi^0\bar\Sigma^0 K^0_S$, and $M_{\bar{p}\pi^0}~\text{vs.}~M_{\Lambda\pi^0}$ for $J/\psi\to\Xi^0\bar\Sigma^- K^+$.
Events are divided into four categories:
the `Signal' describes candidates with both dimensions correctly reconstructed;
the `BKGI' denotes candidates where the $x$-axis mass is true signal and the $y$-axis mass comes from combinatorial background;
the `BKGII' is similar to the `BKGI' with $y$-axis mass being true signal and the $x$-axis mass coming from combinatorial background;
the `BKGIII' encompasses candidates where both masses arise from combinatorial backgrounds, as well as the incorrectly reconstructed events with different final states.

\begin{figure*}[htbp]\centering
  \includegraphics[width=1.0\textwidth]{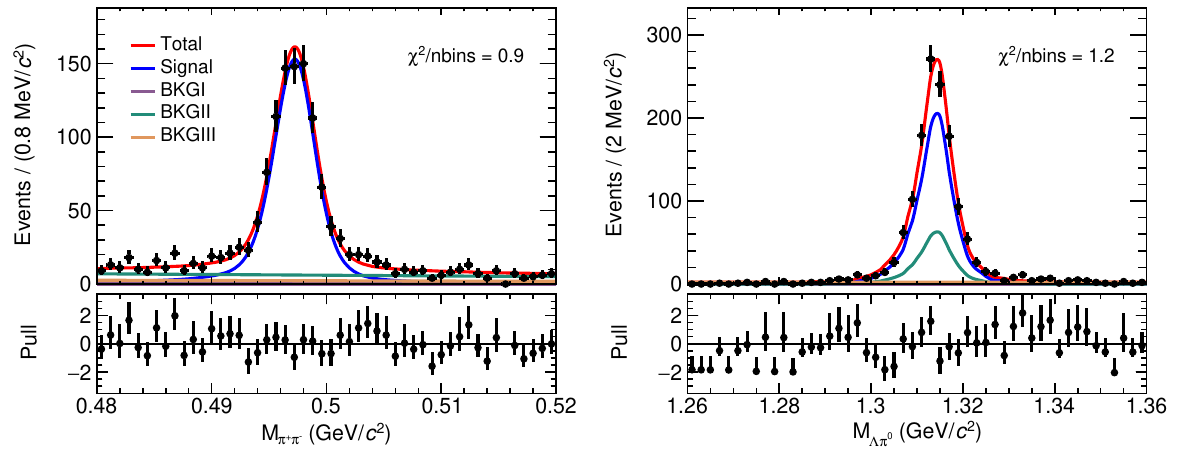}
  \includegraphics[width=1.0\textwidth]{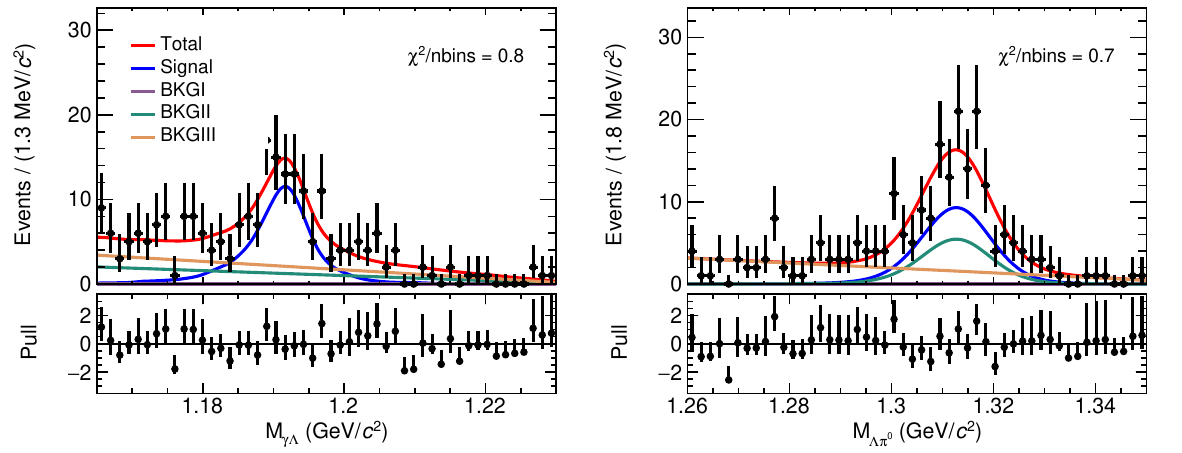}
  \includegraphics[width=1.0\textwidth]{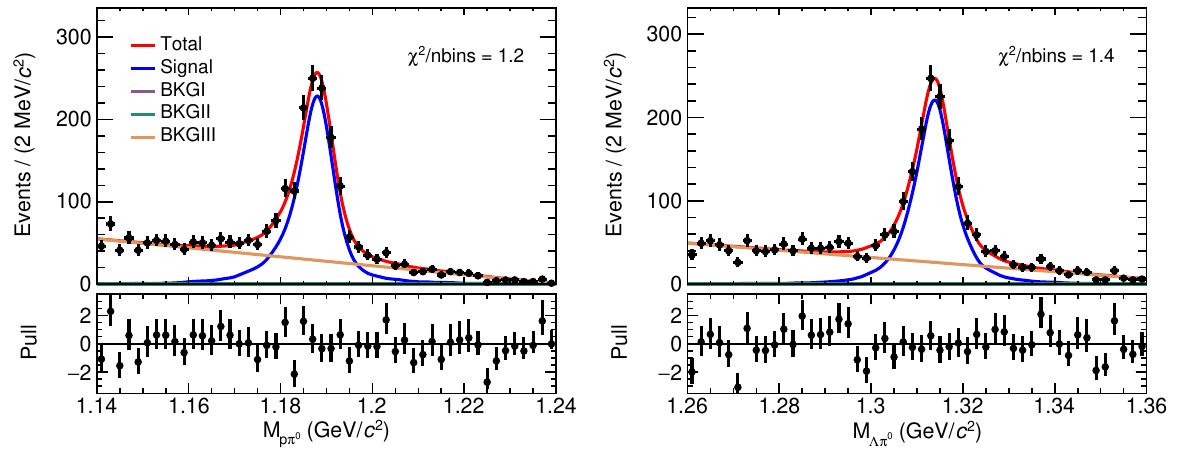}
  \caption{One-dimensional projections of the 2D fit to (top) the $M_{\pi^+\pi^-}~\text{vs}~M_{\Lambda\pi^0}$ distribution of $J/\psi\to\Xi^0\bar\Lambda K^0_S$, (middle) the $M_{\gamma\bar\Lambda}~\text{vs}~M_{\Lambda\pi^0}$ distribution of $J/\psi\to\Xi^0\bar\Sigma^0 K^0_S$, and (bottom) the $M_{\bar{p}\pi^0}~\text{vs}~M_{\Lambda\pi^0}$ distribution of $J/\psi\to\Xi^0\bar\Sigma^- K^+$. The dots with error bars are data events, the red solid curves are the total fit results, while the blue curves are the signal contributions of the fit and other curves represent the different background contributions. For each projection, the $\chi^2$/nbins are provided, with $\chi^2$ being calculated from the difference between the binned data points and the total fit projection, and nbins representing the number of bins.}
  \label{fig:fit}
\end{figure*}

The probability density functions (PDFs) of Signal, BKGI, BKGII, and BKGIII are constructed as:
\begin{itemize}
  \item Signal: ${\mathcal S}_x \times {\mathcal S}_y$,
  \item BKGI: ${\mathcal S}_x \times {\mathcal A}_y$,
  \item BKGII: ${\mathcal A}_x \times {\mathcal S}_y$,
  \item BKGIII: ${\mathcal A}_x \times {\mathcal A}_y$.
\end{itemize}
Here, $x$ and $y$ correspond to the two dimensions of the 2D fit. The ${\mathcal S}_x$ and ${\mathcal S}_y$ are the signal shapes derived from signal MC simulation and convolved with a Gaussian resolution function with free parameters, 
while ${\mathcal A}_x$ and ${\mathcal A}_y$ are the second order polynomials which characterize the combinatorial background shape in the corresponding invariant mass distributions.
The signal yields are $982\pm 37$, $91\pm 13$, and $1284\pm 39$ for $J/\psi\to\Xi^0\bar\Lambda K^0_S$, $J/\psi\to\Xi^0\bar\Sigma^0 K^0_S$, and $J/\psi\to\Xi^0\bar\Sigma^- K^+$, respectively. 
The fit projections are illustrated in Fig.~\ref{fig:fit}. 
The statistical significance for each channel is found to exceed $5\sigma$, evaluated by comparing the change in negative log-likelihood between fits with and without the signal component, taking into account the difference in the number of degrees of freedom.

\subsection{Detection efficiencies}

\begin{figure*}[htbp]\centering
  \includegraphics[keepaspectratio=true,width=0.495\textwidth,angle=0]{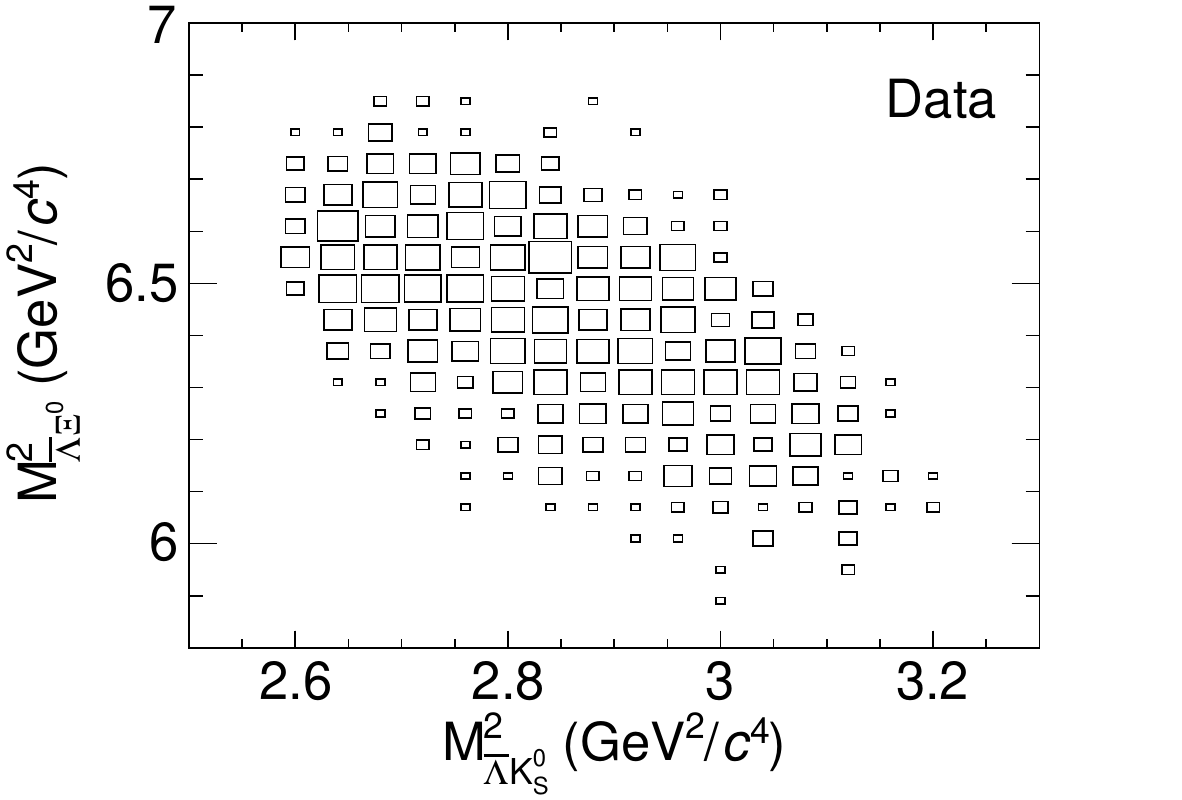}
  \includegraphics[keepaspectratio=true,width=0.495\textwidth,angle=0]{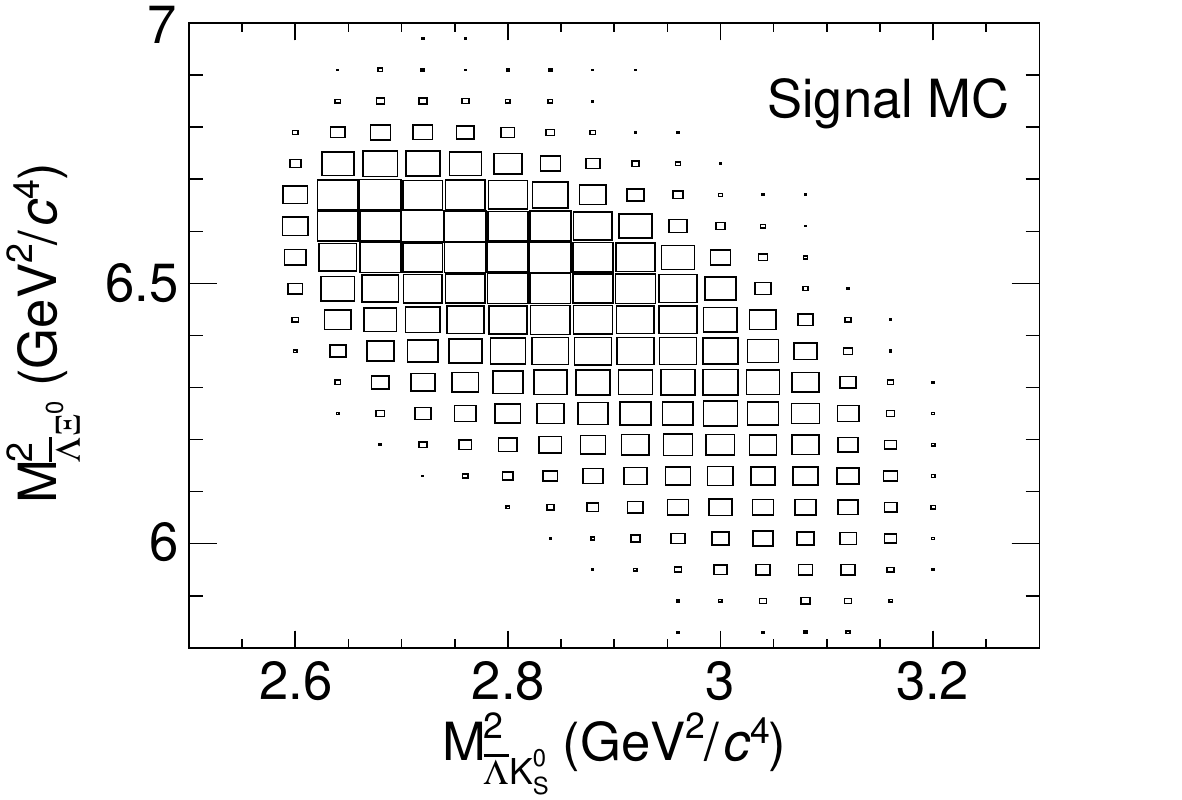}
  \includegraphics[keepaspectratio=true,width=0.495\textwidth,angle=0]{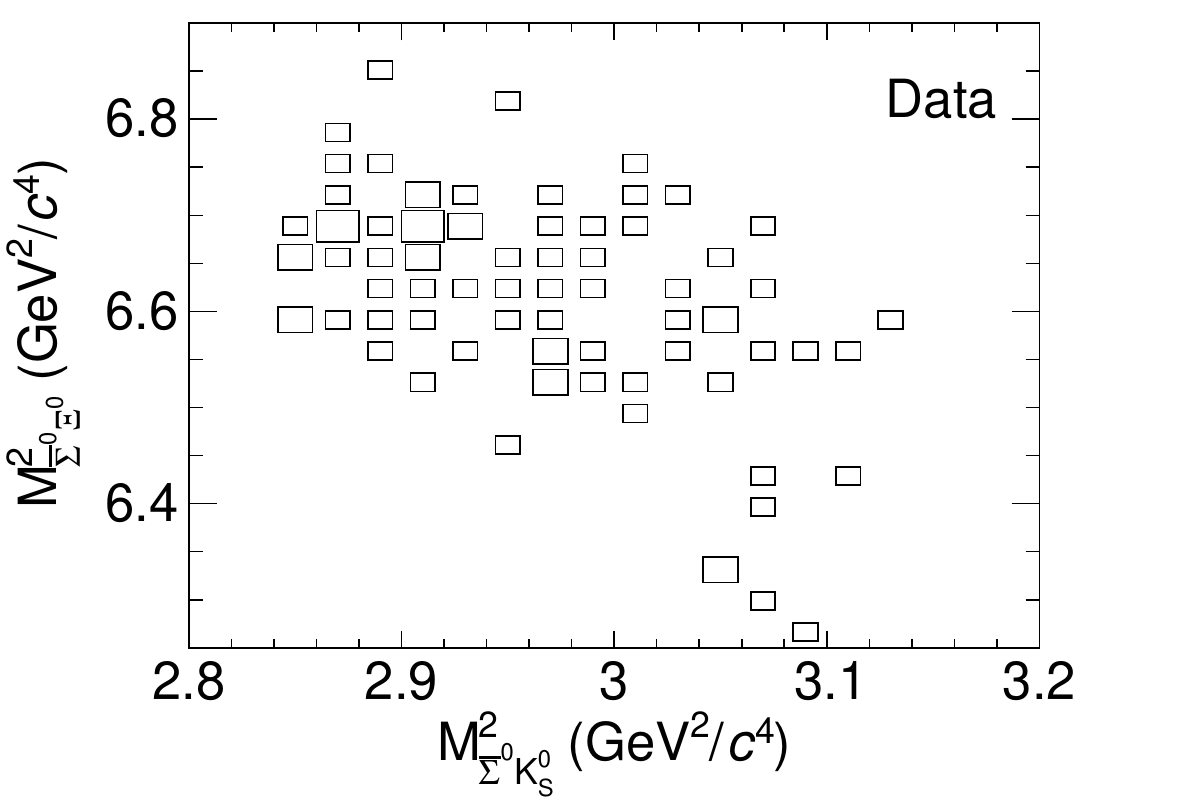}
  \includegraphics[keepaspectratio=true,width=0.495\textwidth,angle=0]{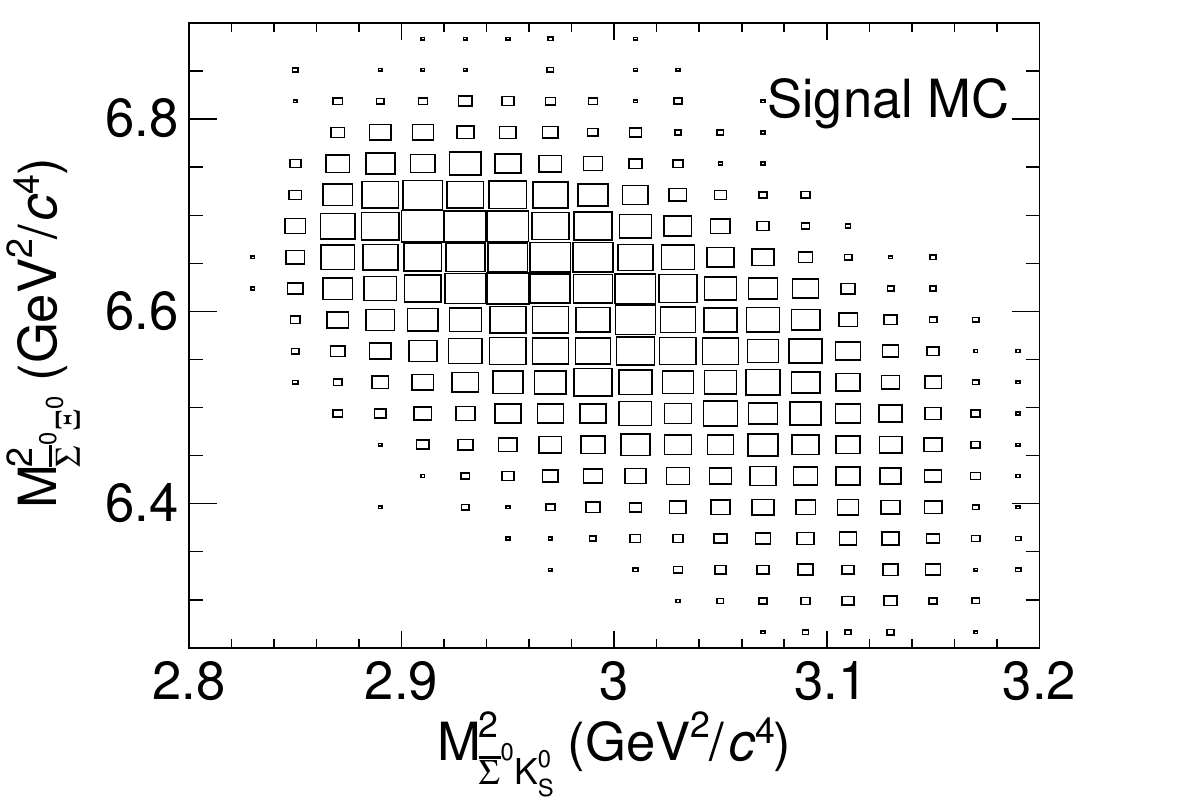}
  \includegraphics[keepaspectratio=true,width=0.495\textwidth,angle=0]{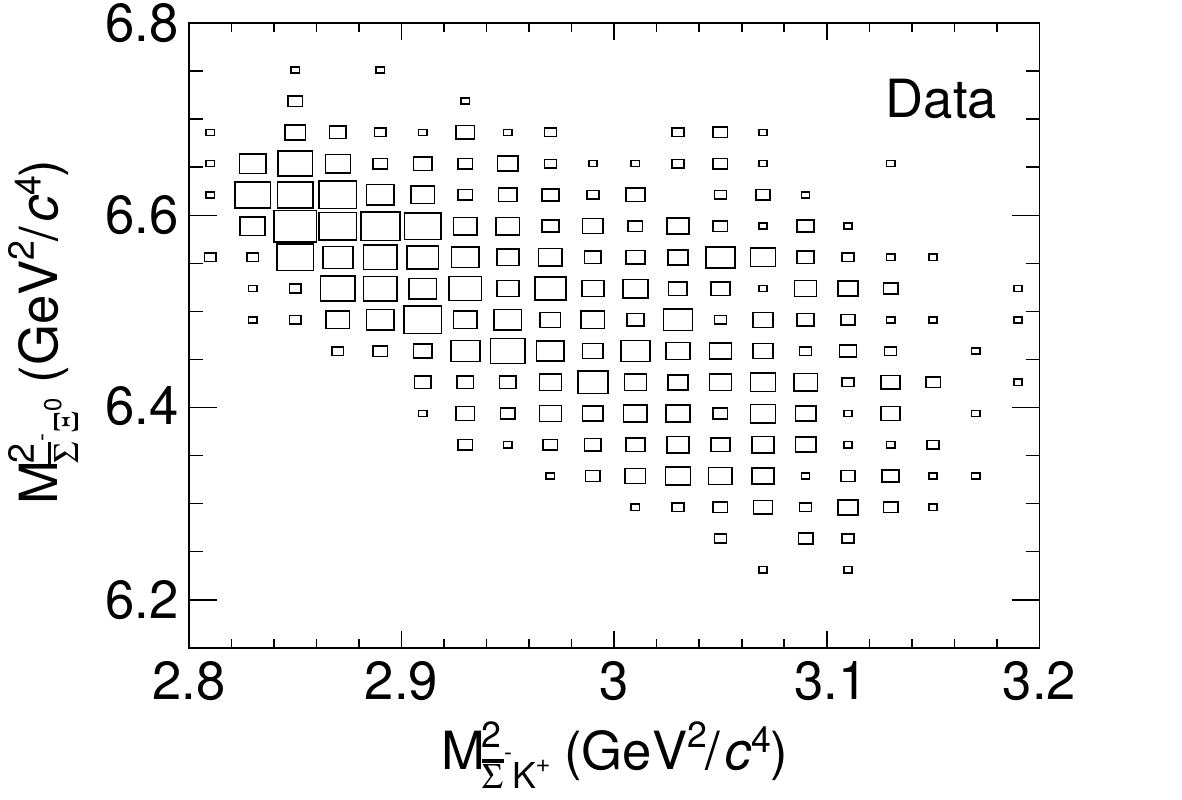}
  \includegraphics[keepaspectratio=true,width=0.495\textwidth,angle=0]{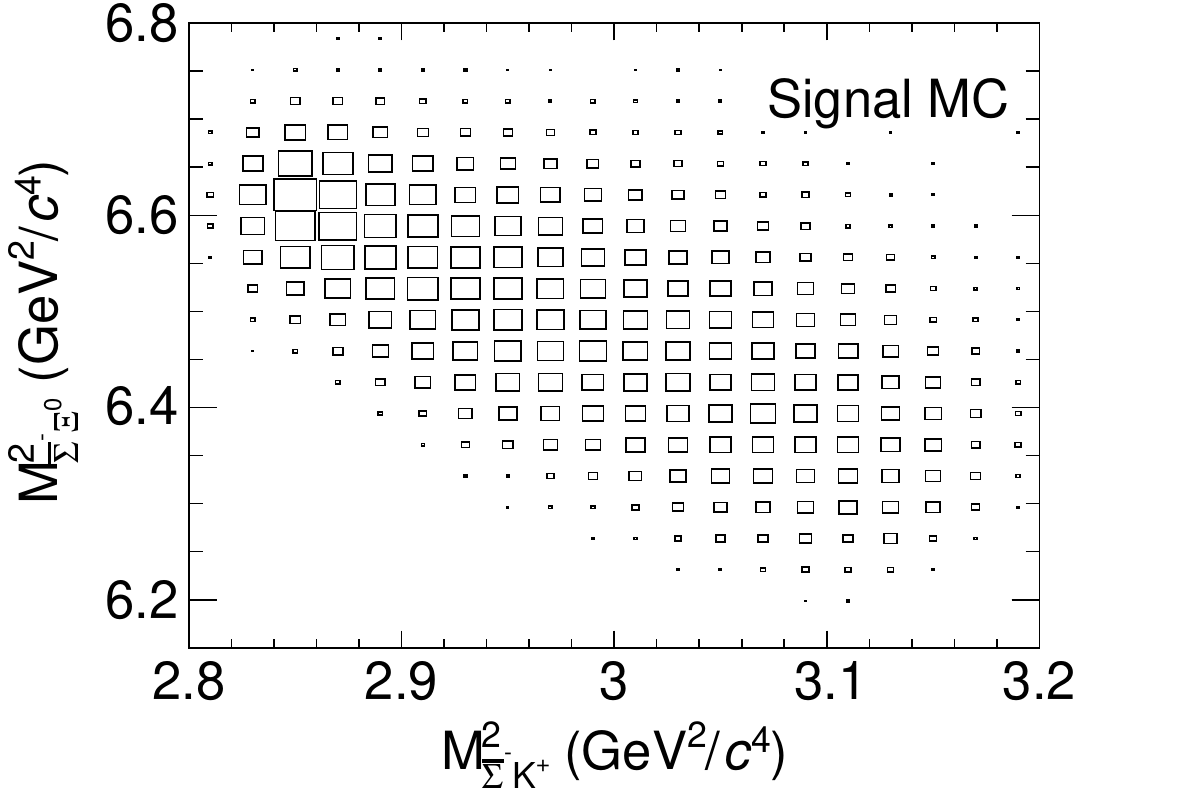}
  \caption{Dalitz plots of (top) $M^2_{\bar \Lambda \Xi^0}$ vs $ M^2_{\bar \Lambda K_S^0}$ of the candidates for $J/\psi\to \Xi^0\bar\Lambda K^0_S$, (middle) $M^2_{\bar \Sigma^0 \Xi^0}$ vs $ M^2_{\bar \Sigma^0 K_S^0}$ of the candidates for $J/\psi\to \Xi^0\bar\Sigma^0 K^0_S$ and (bottom) $M^2_{\bar\Sigma^-\Xi^0}$ vs $ M^2_{\bar\Sigma^- K^+}$ of the candidates for $J/\psi\to\Xi^0\bar\Sigma^- K^+$.}
  \label{fig:dalitz}
\end{figure*}

\begin{figure*}[htbp]\centering
  \includegraphics[width=0.32\textwidth]{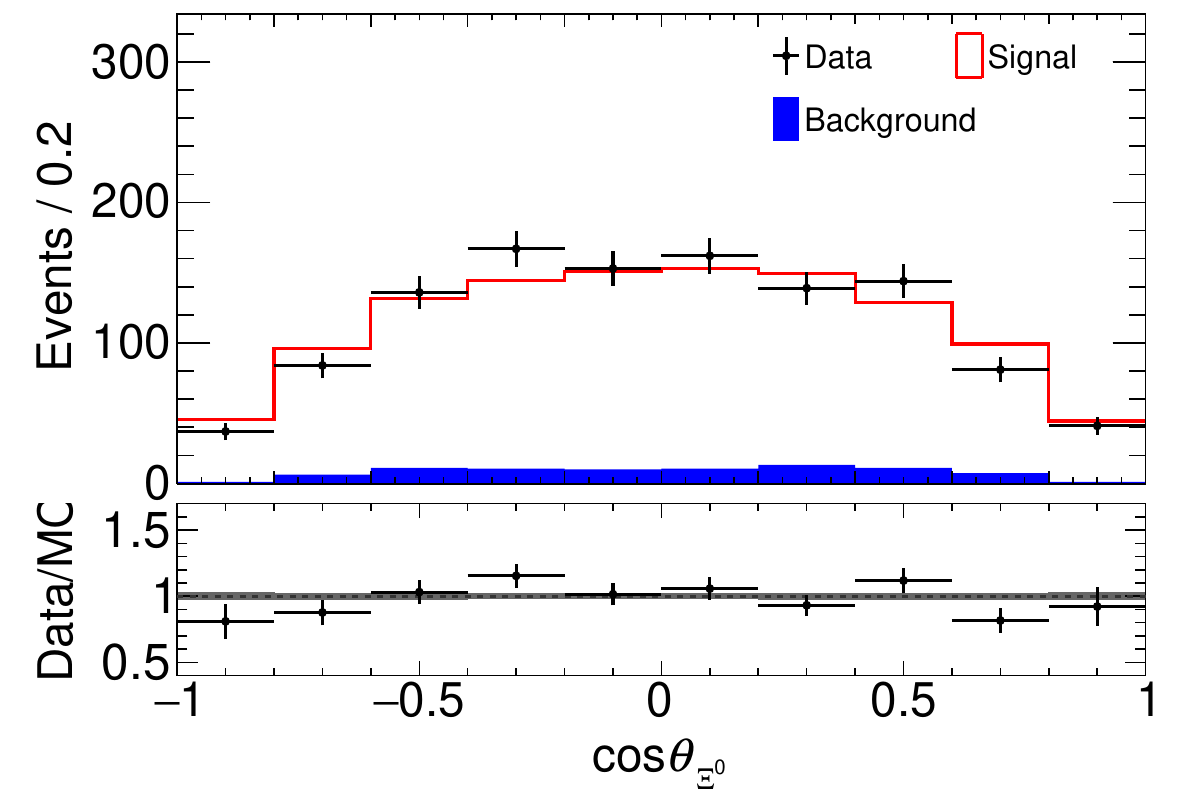}
  \includegraphics[width=0.32\textwidth]{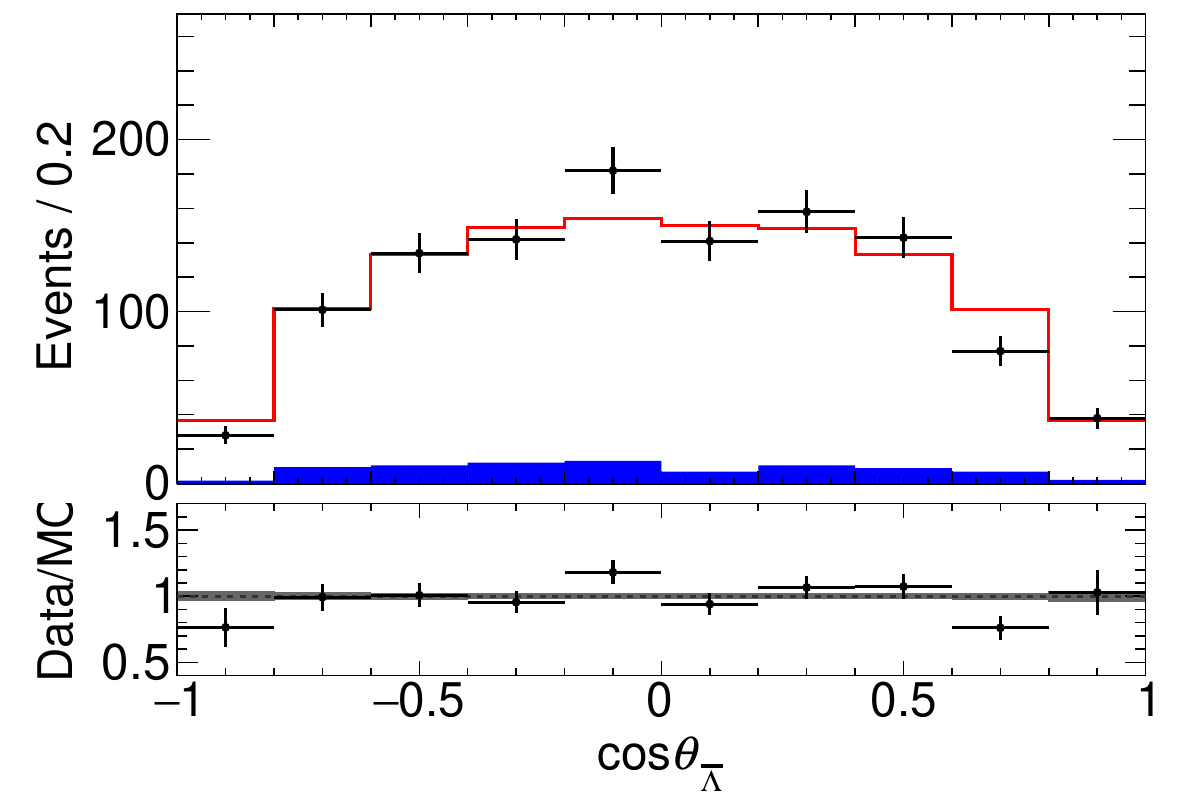}
  \includegraphics[width=0.32\textwidth]{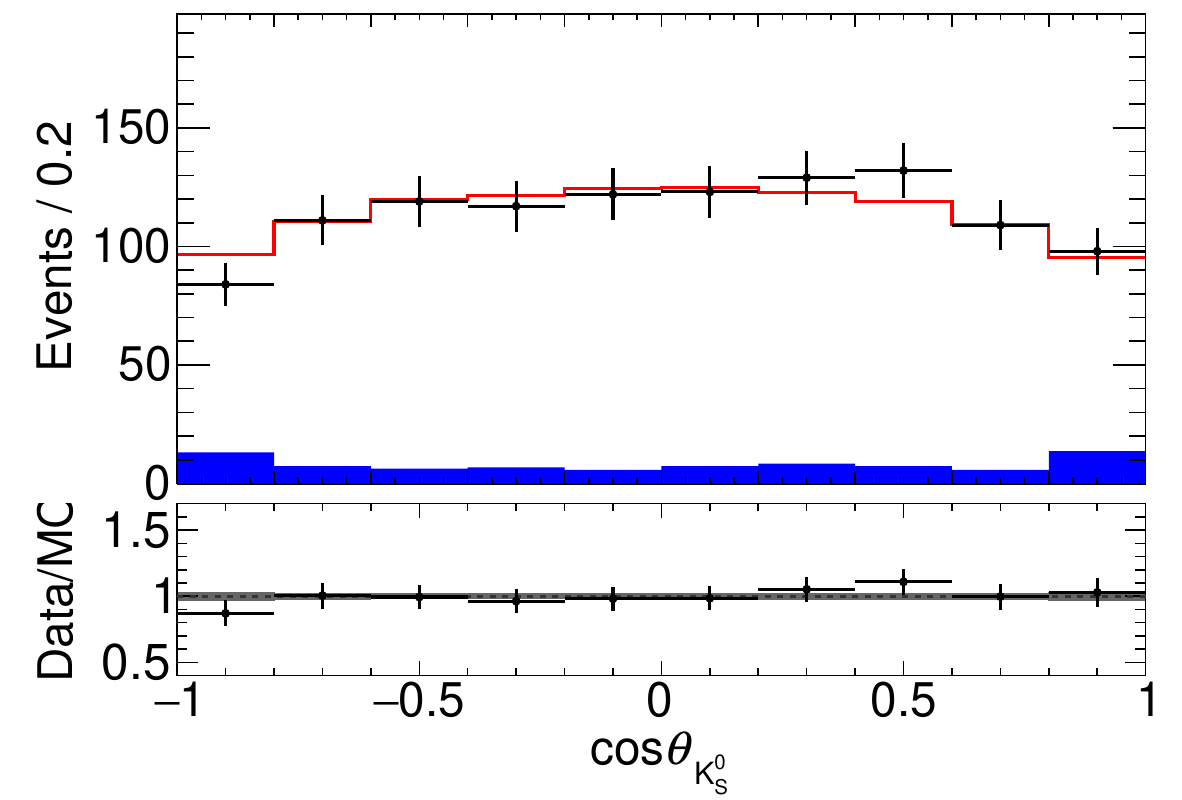}
  \includegraphics[width=0.32\textwidth]{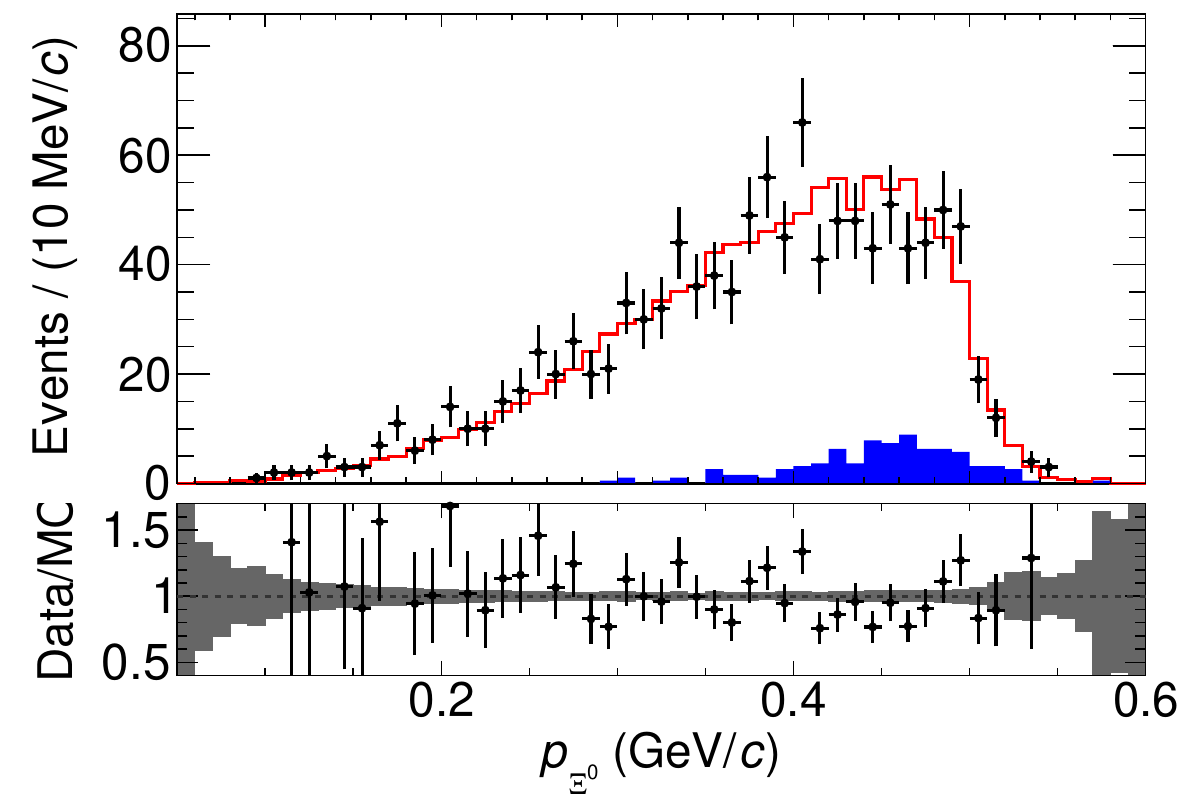}
  \includegraphics[width=0.32\textwidth]{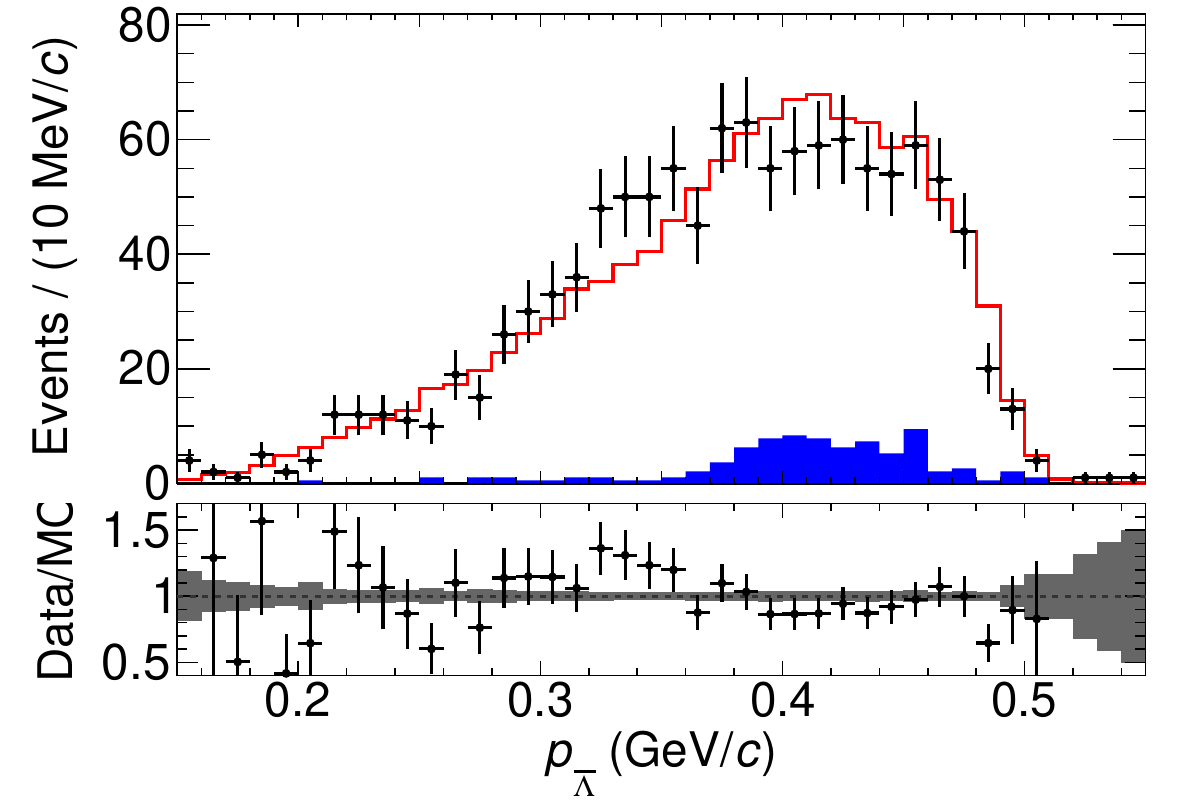}
  \includegraphics[width=0.32\textwidth]{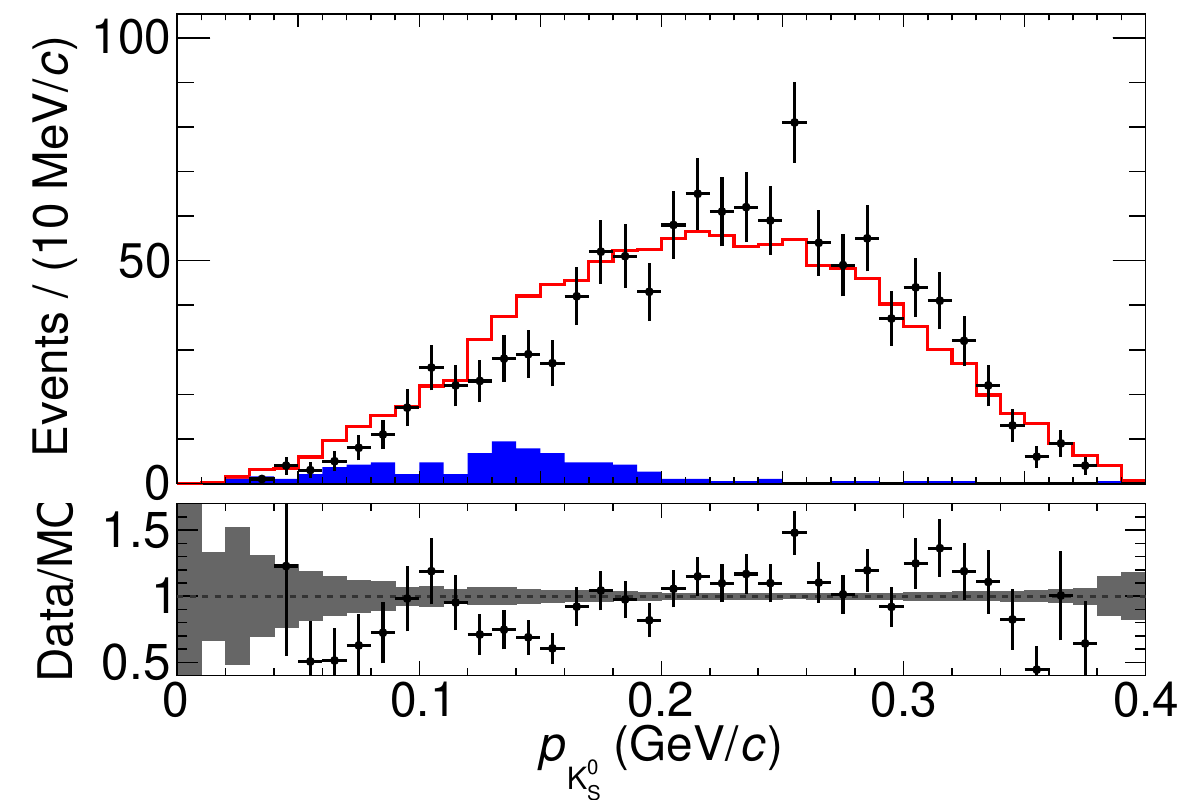}
  \includegraphics[width=0.32\textwidth]{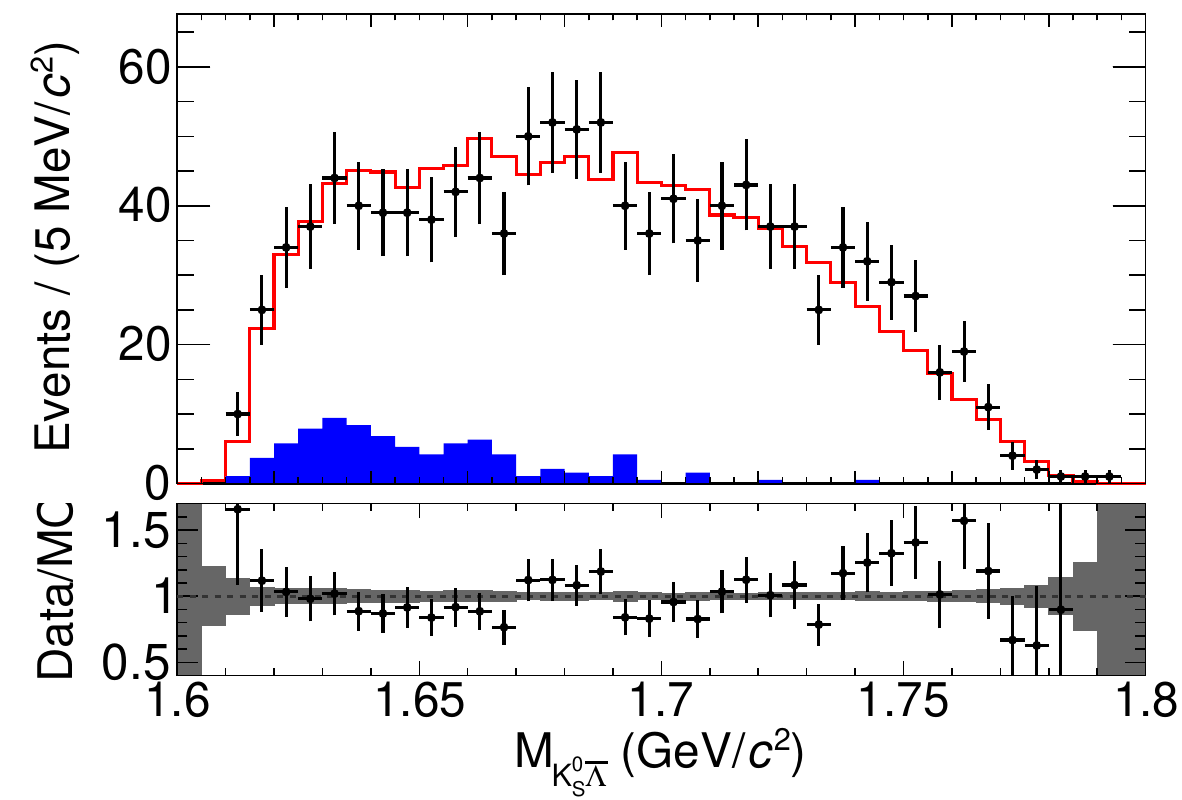}
  \includegraphics[width=0.32\textwidth]{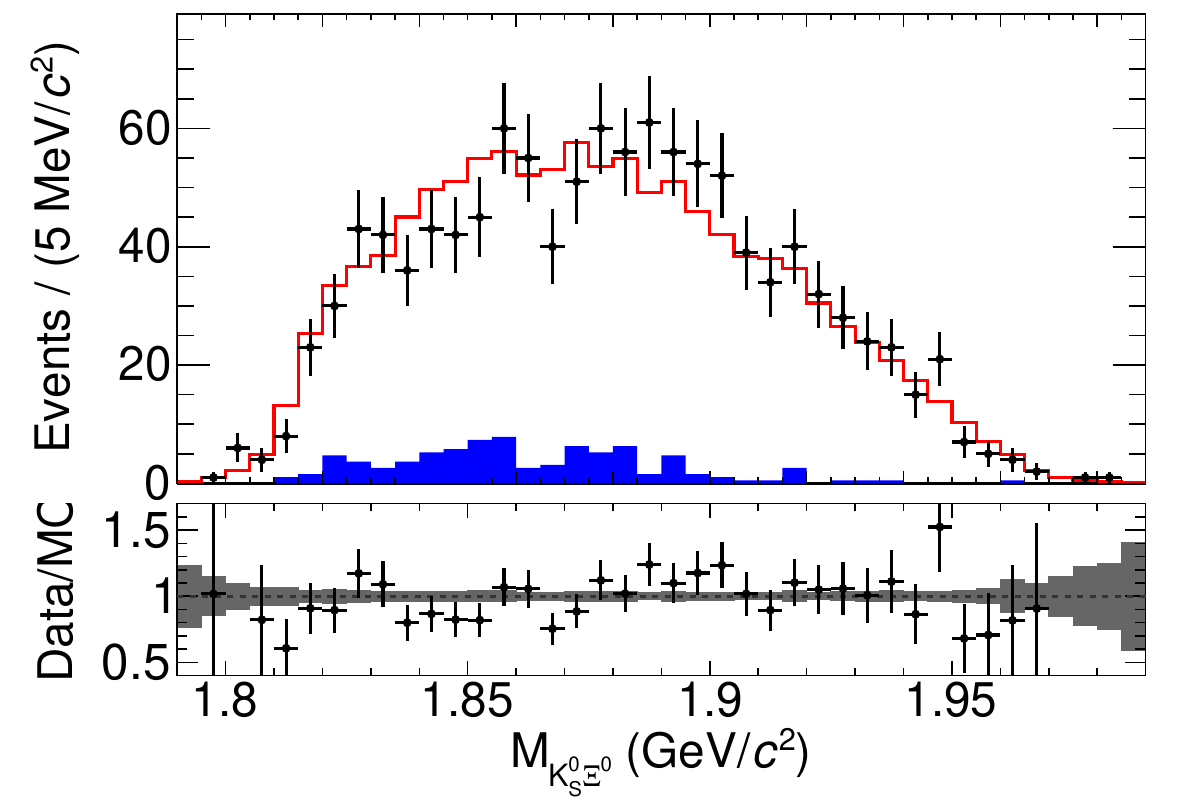}
  \includegraphics[width=0.32\textwidth]{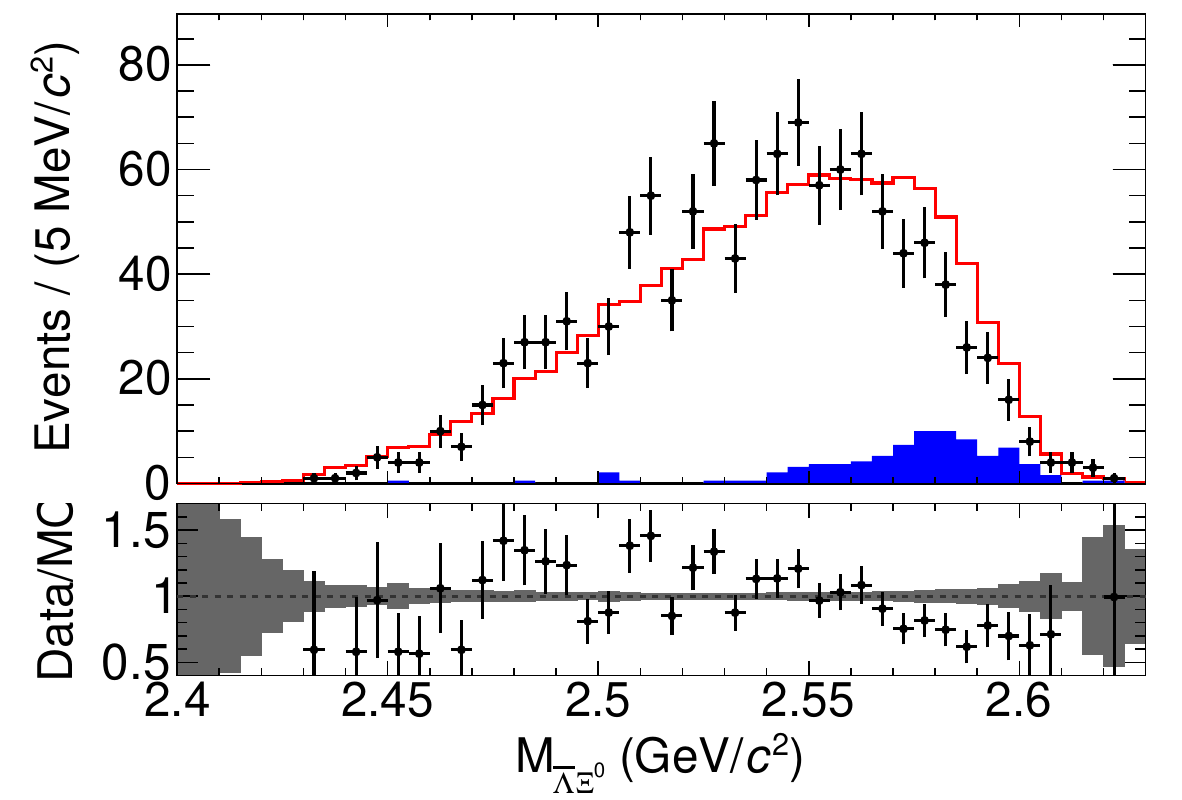}
  \caption{Distributions of the momenta, cosines of polar angles of the $\Xi^0$, $\bar\Lambda$ and $K_S^0$ candidates as well as two-body mass distributions for the $J/\psi\to\Xi^0\bar\Lambda K^0_S$ candidates. The black points with error bars are data, the solid red lines show the signal MC simulation which is scaled to the total number of events of data, and the blue solid-filled histograms are the background contribution of the inclusive MC sample. The bottom panels show the data and MC comparison, where the error bands indicate the MC statistical uncertainty only.}
  \label{fig:somefig1}
\end{figure*}

\begin{figure*}[htbp]\centering
  \includegraphics[width=0.32\textwidth]{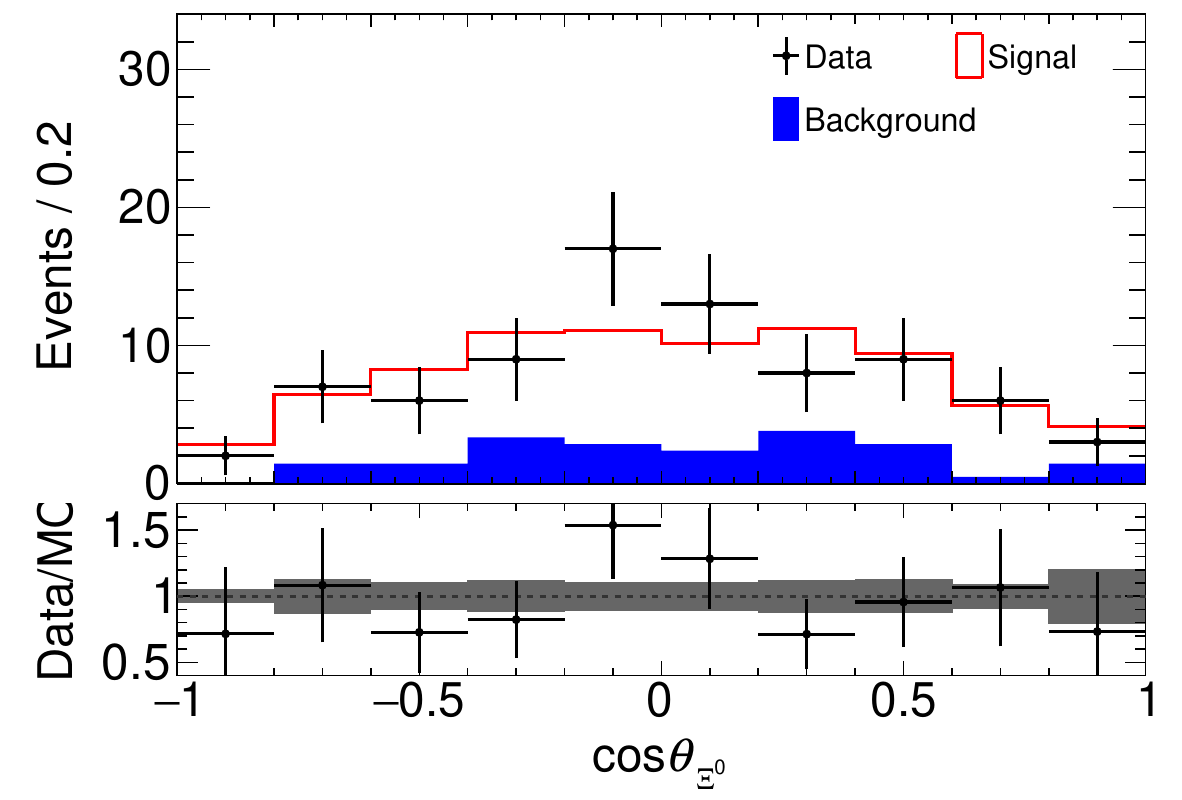}
  \includegraphics[width=0.32\textwidth]{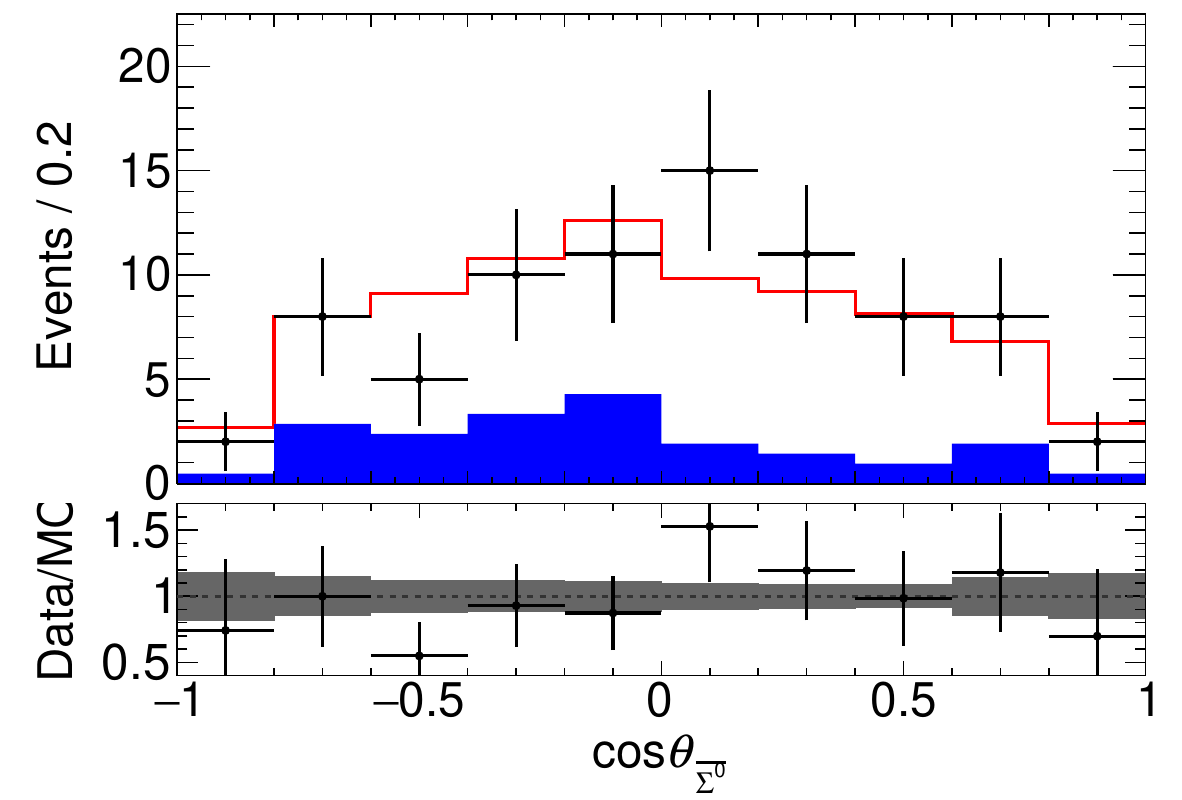}
  \includegraphics[width=0.32\textwidth]{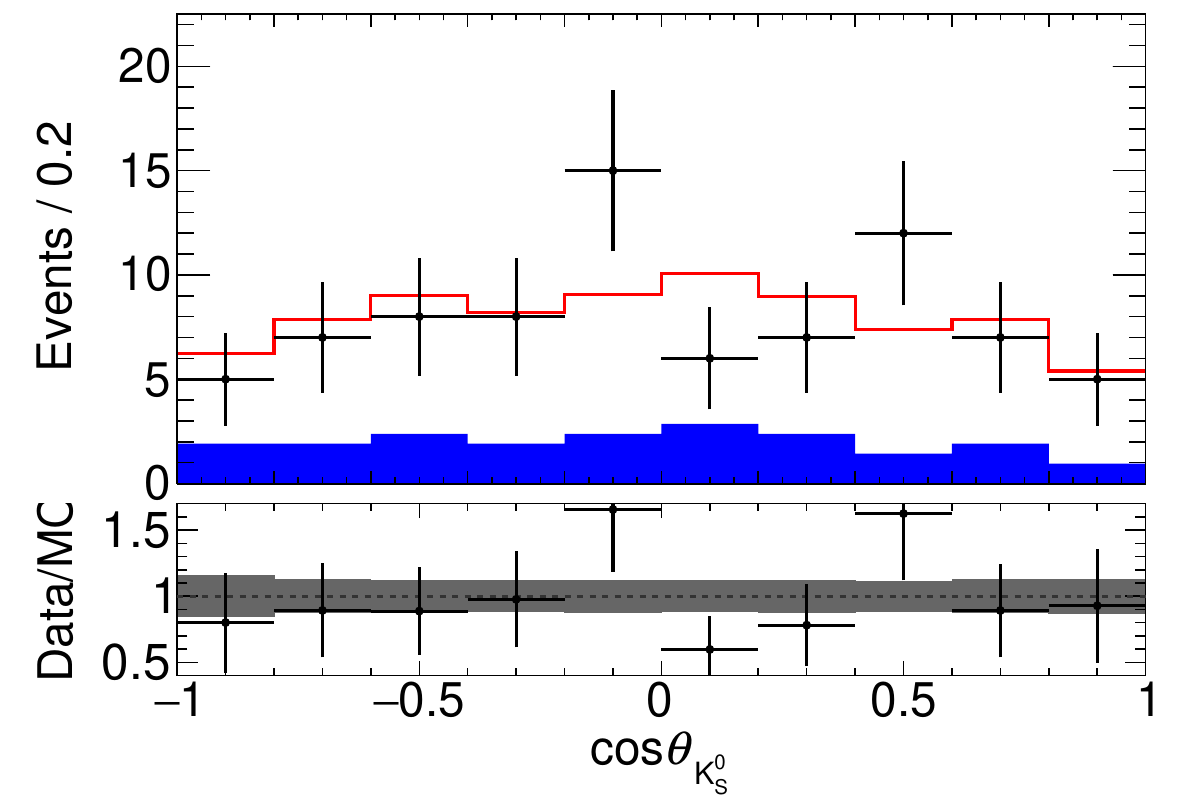}
  \includegraphics[width=0.32\textwidth]{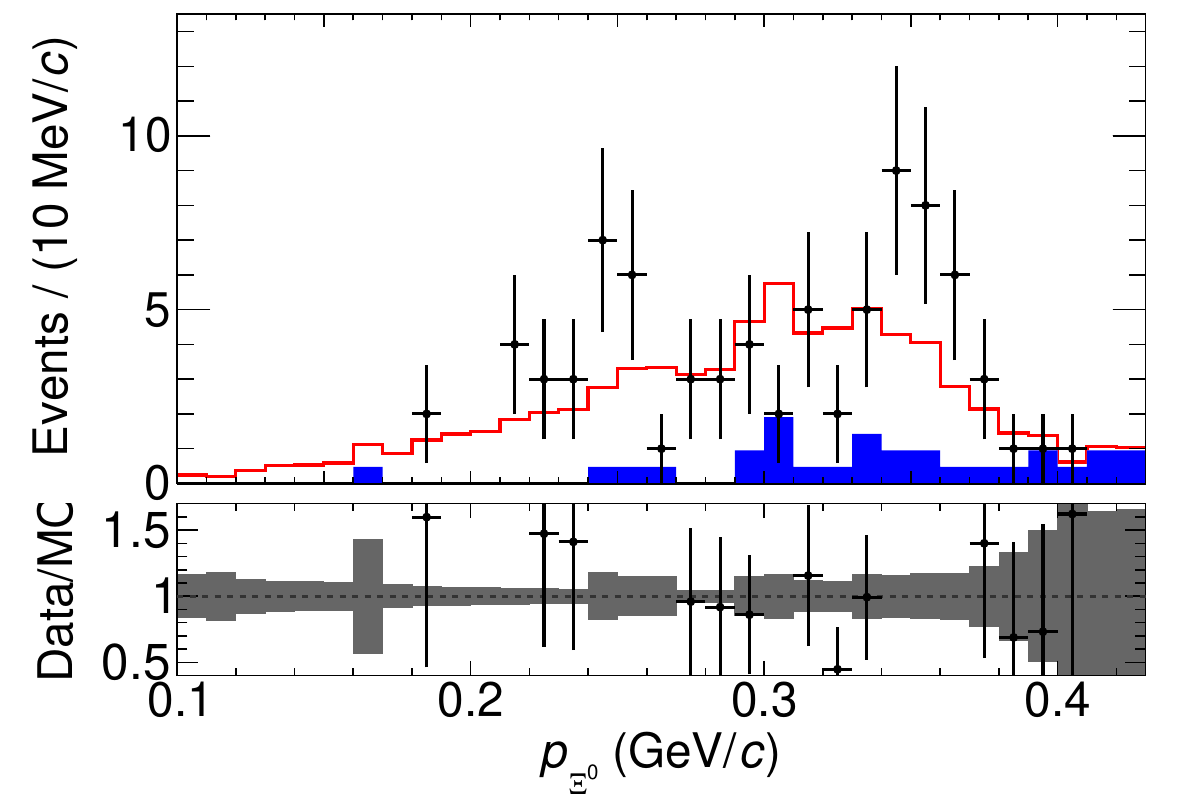}
  \includegraphics[width=0.32\textwidth]{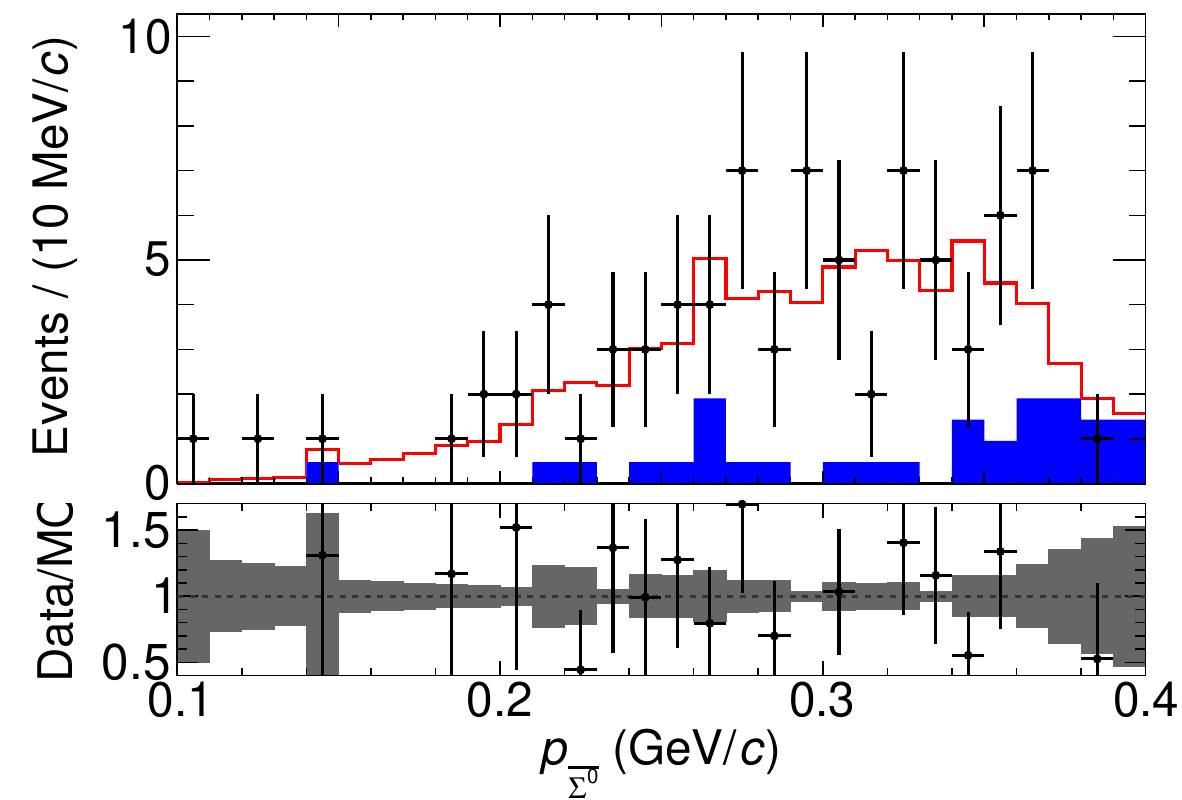}
  \includegraphics[width=0.32\textwidth]{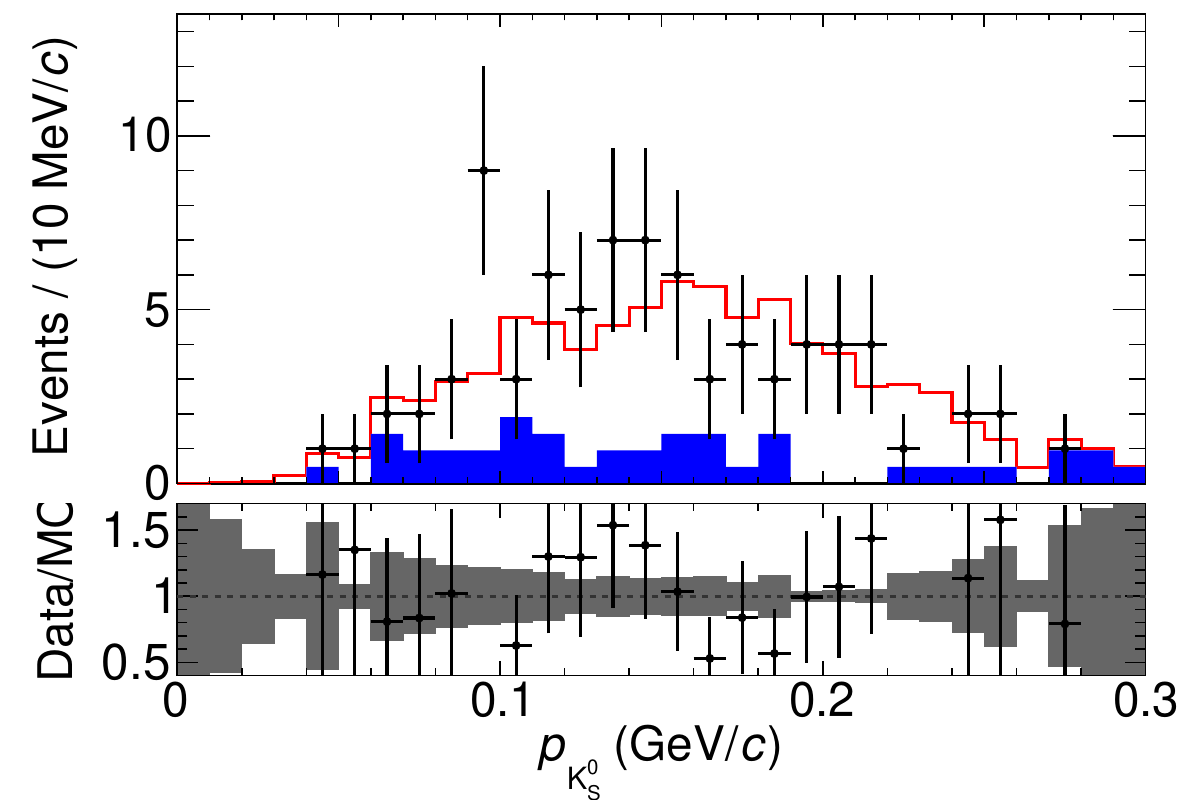}
  \includegraphics[width=0.32\textwidth]{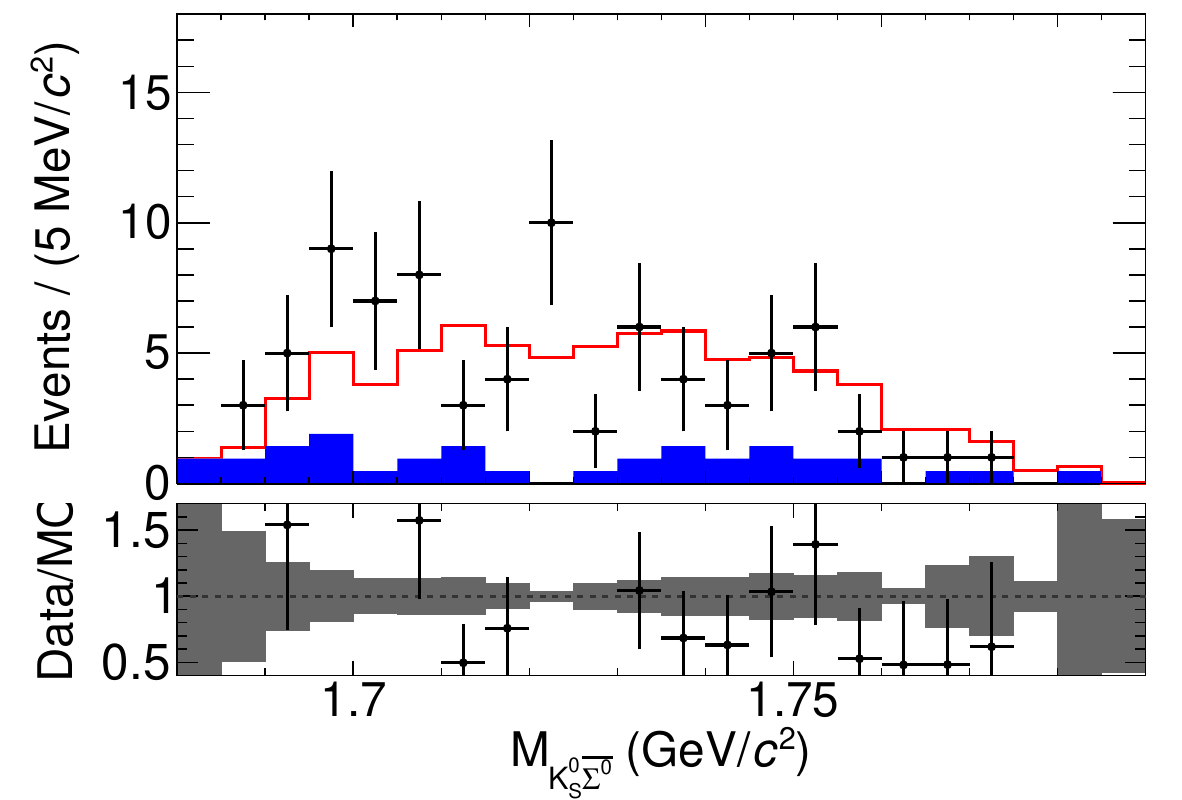}
  \includegraphics[width=0.32\textwidth]{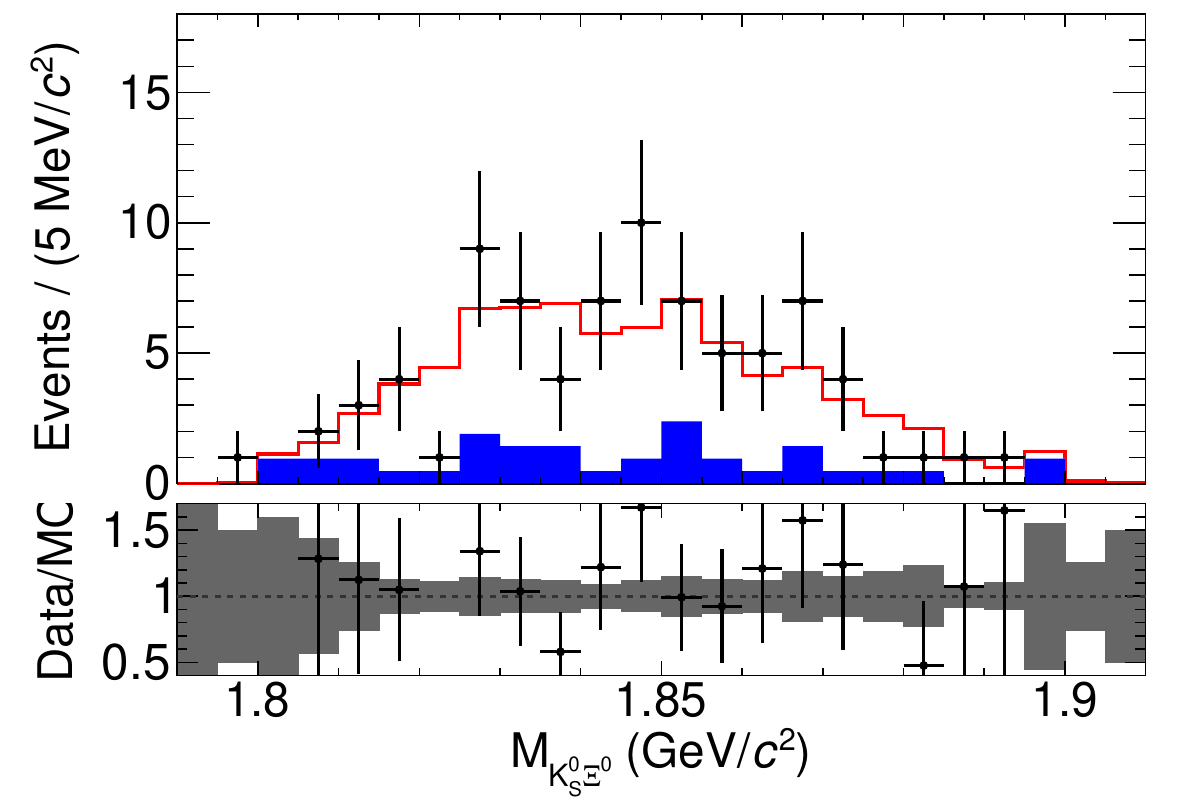}
  \includegraphics[width=0.32\textwidth]{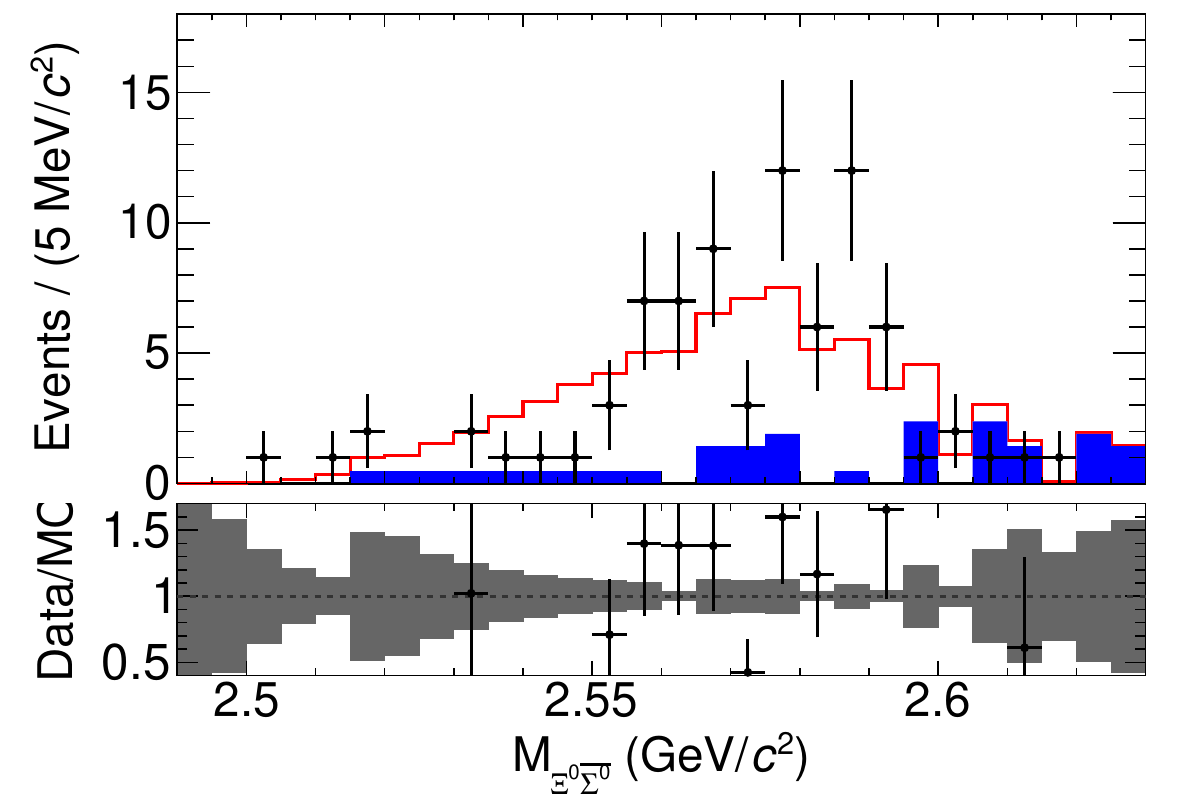}
  \caption{Distributions of the momenta, cosines of polar angles of the $\Xi^0$, $\bar\Sigma^0$, and $K_S^0$ candidates as well as two-body mass distributions for the $J/\psi\to \Xi^0\bar\Sigma^0 K^0_S$ candidates. The black points with error bars are data, the solid red lines show the signal MC simulation which is scaled to the total number of events of data, and the blue solid-filled histograms are the background contribution of the inclusive MC sample. The bottom panels show the data and MC comparison, where the error bands indicate the MC statistical uncertainty only.}
  \label{fig:somefig2}
\end{figure*}

\begin{figure*}[htbp]\centering
  \includegraphics[width=0.32\textwidth]{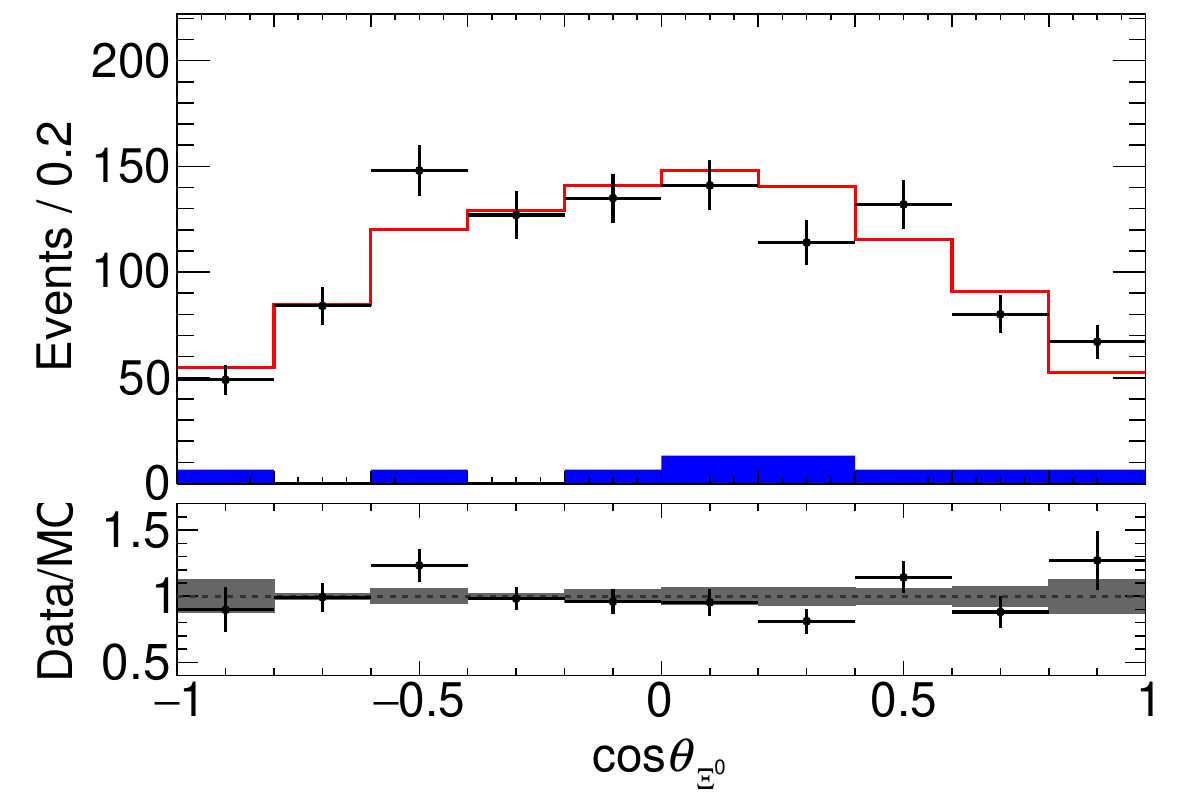}
  \includegraphics[width=0.32\textwidth]{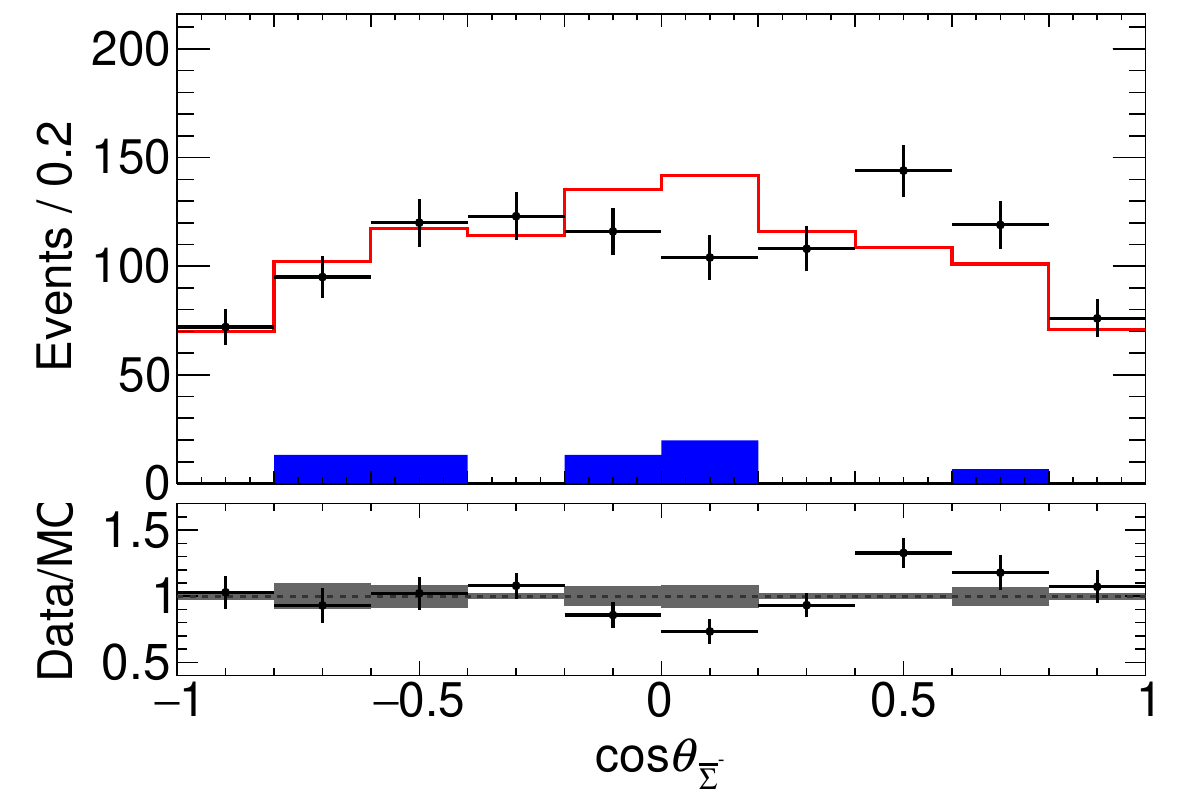}
  \includegraphics[width=0.32\textwidth]{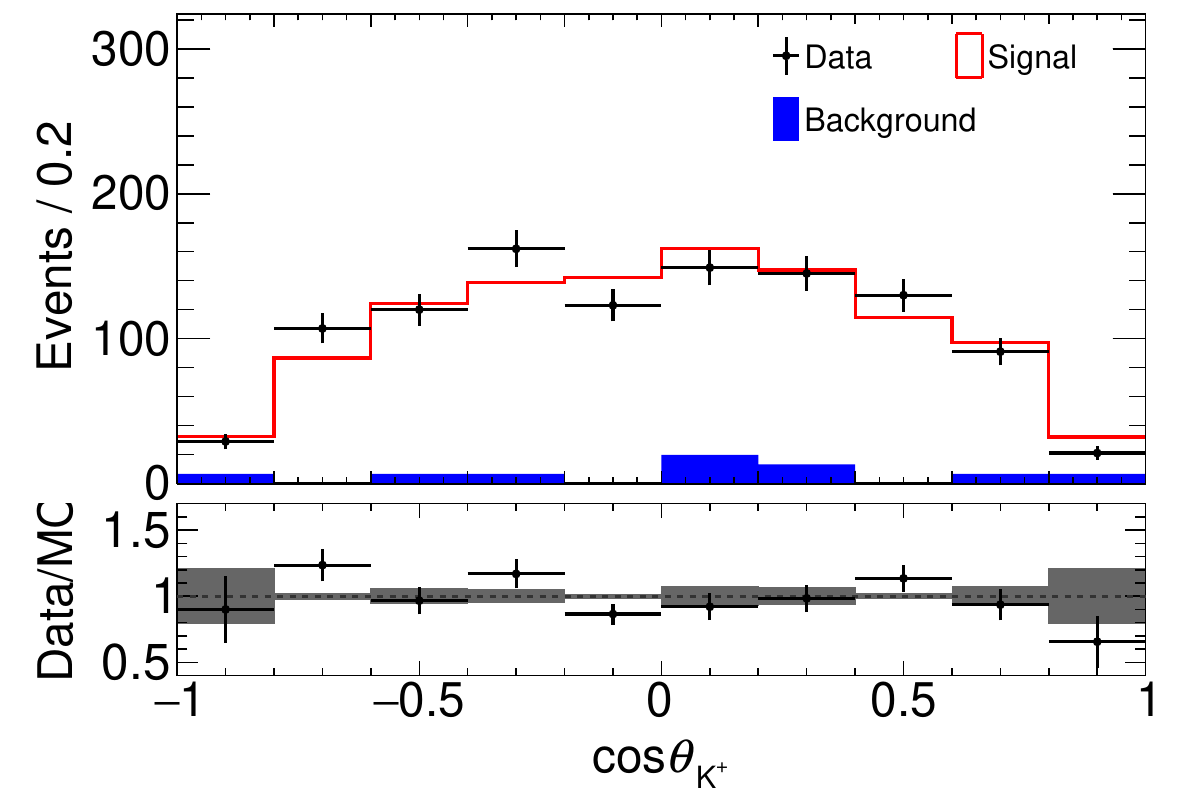}
  \includegraphics[width=0.32\textwidth]{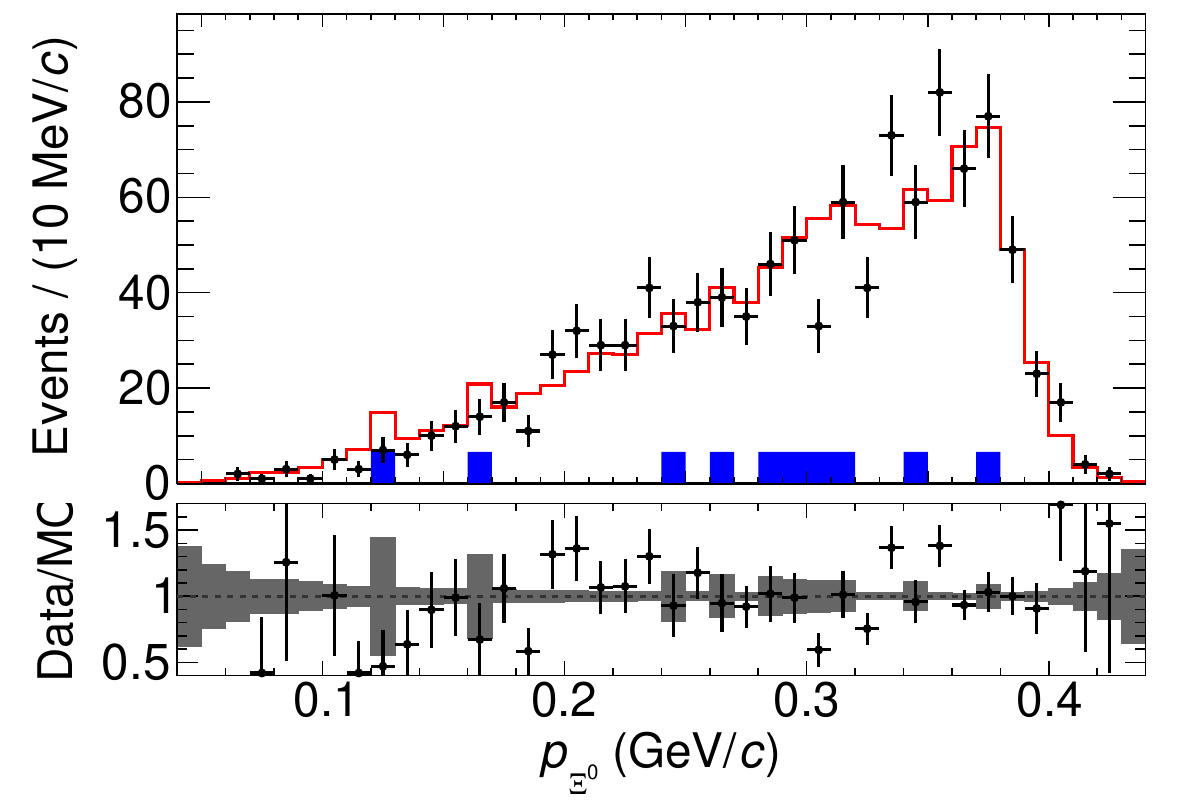}
  \includegraphics[width=0.32\textwidth]{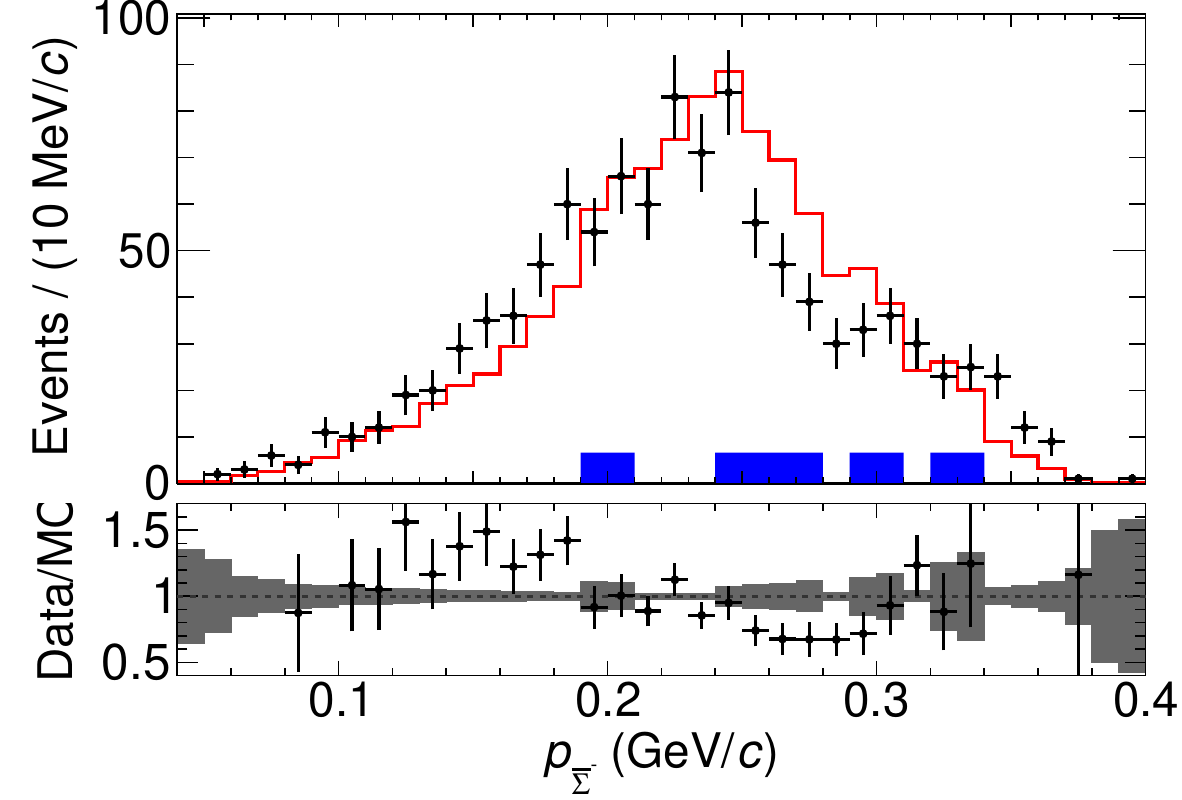}
  \includegraphics[width=0.32\textwidth]{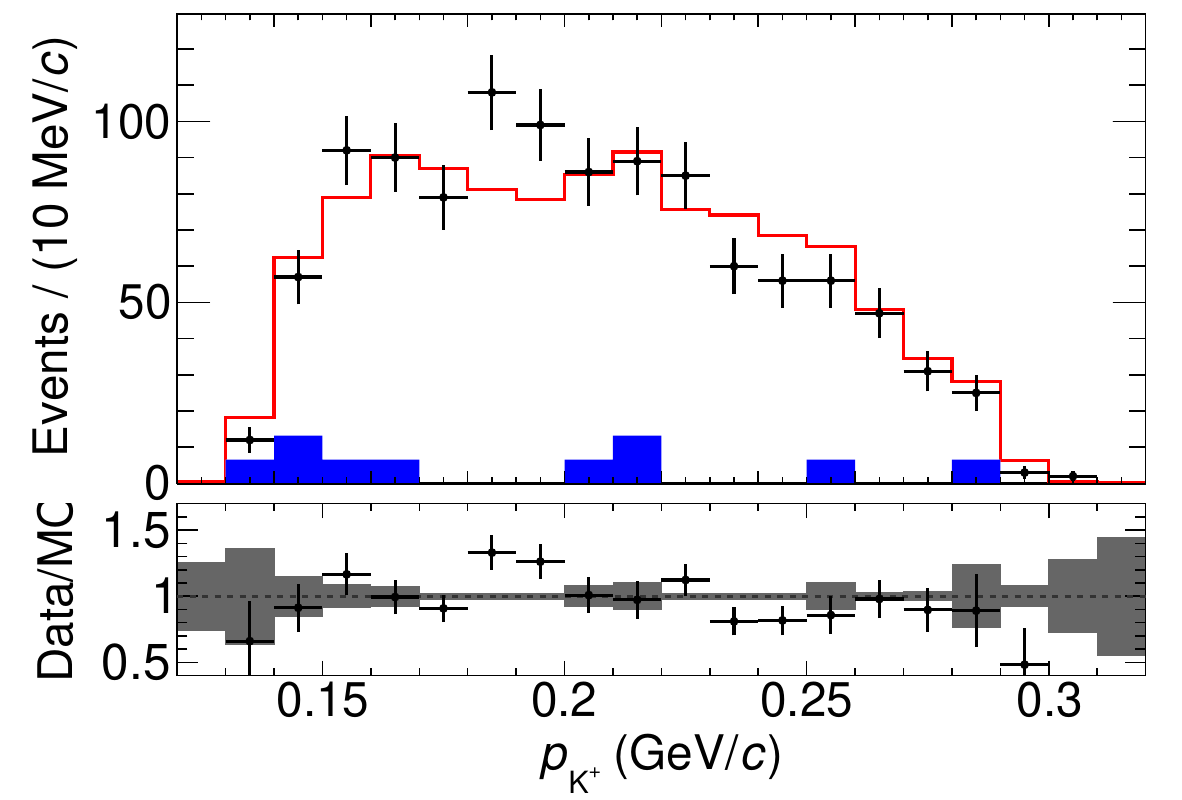}
  \includegraphics[width=0.32\textwidth]{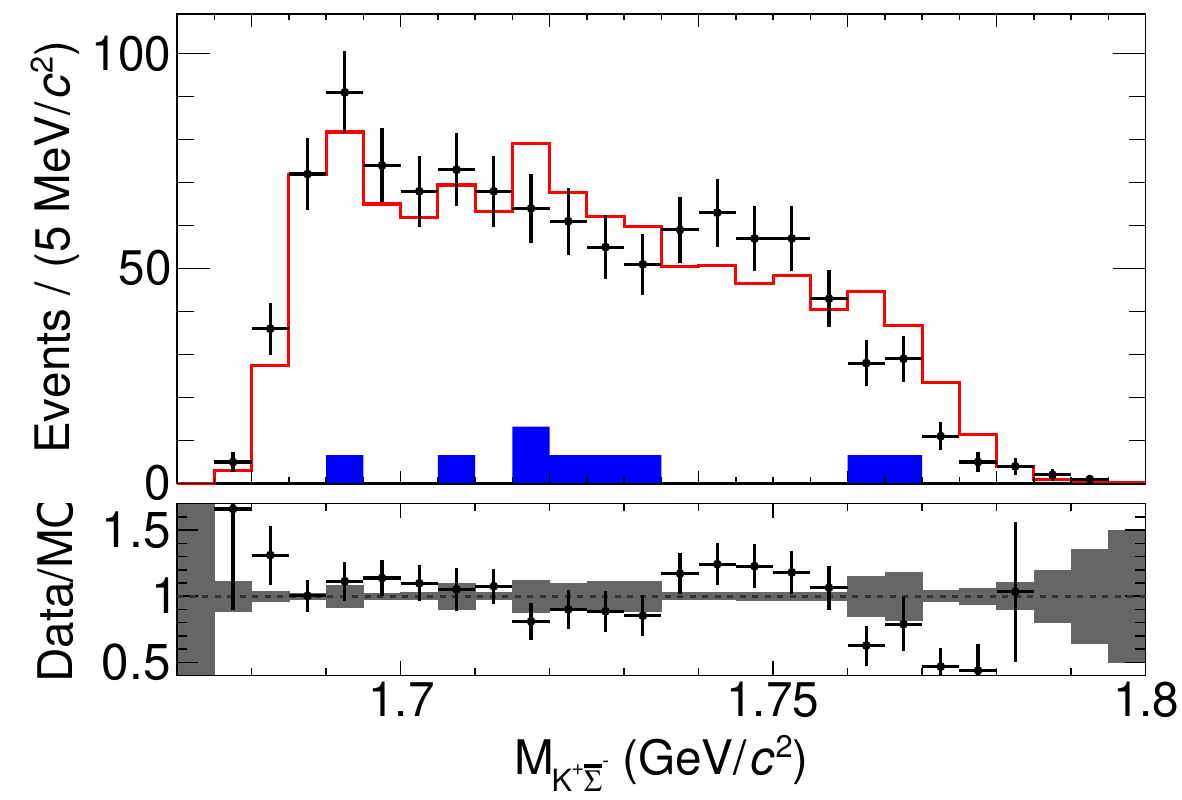}
  \includegraphics[width=0.32\textwidth]{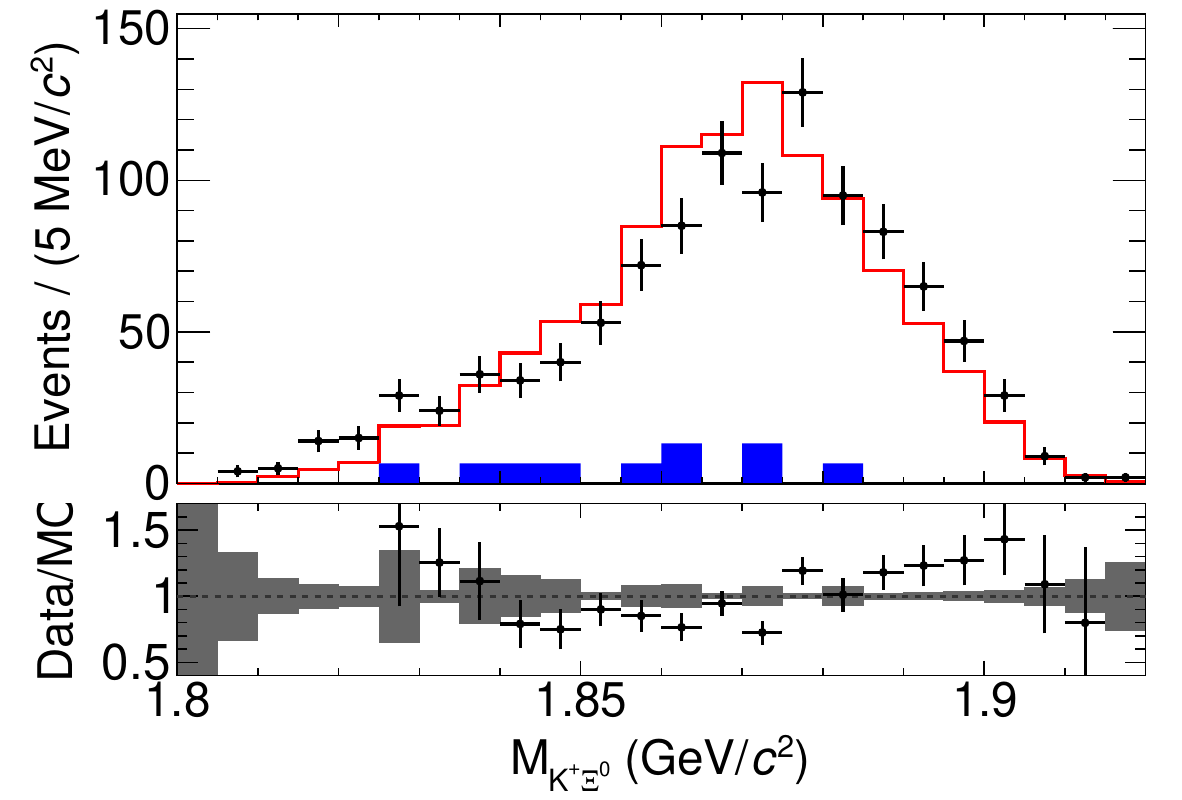}
  \includegraphics[width=0.32\textwidth]{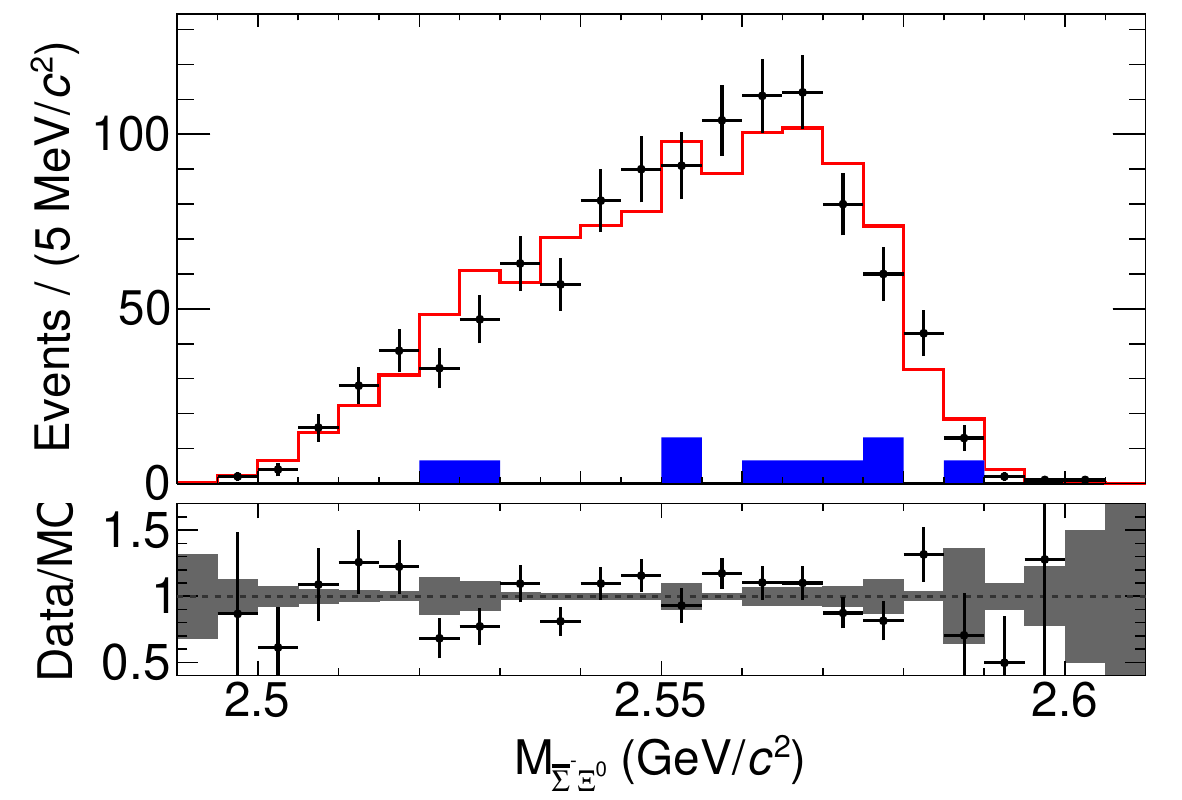}
  \caption{Distributions of the momenta, cosines of polar angles of the $\Xi^0$, $\bar\Sigma^-$, and $K^+$ candidates as well as two-body mass distributions for the $J/\psi\to \Xi^0\bar\Sigma^- K^+$ candidates. The black points with error bars are data, the solid red lines show the signal MC simulation which is scaled to the total number of events of data, and the blue solid-filled histograms are the background contribution of the inclusive MC sample. The bottom panels show the data and MC comparison, where the error bands indicate the MC statistical uncertainty only.}
  \label{fig:somefig3}
\end{figure*}

The detection efficiencies of the $J/\psi\to\Xi^0\bar\Lambda K^0_S$, $J/\psi\to\Xi^0\bar\Sigma^0 K^0_S$, and $J/\psi\to\Xi^0\bar\Sigma^- K^+$ decays are evaluated by analyzing the signal MC samples. Figure~\ref{fig:dalitz} shows the Dalitz plots of the candidates selected in data and signal MC samples, with additional requirements of $|M_{\pi^+\pi^-}-m_{K_S^0}|<3$ MeV/$c^2$ and $|M_{\Lambda\pi^0}-m_{\Xi^0}|<30$ MeV/$c^2$ for $J/\psi\to\Xi^0\bar\Lambda K^0_S$, $|M_{\gamma\bar\Lambda}-m_{\Sigma^0}|<6$ MeV/$c^2$ and $|M_{\Lambda\pi^0}-m_{\Xi^0}|<15$ MeV/$c^2$ for $J/\psi\to\Xi^0\bar\Sigma^0 K^0_S$, as well as $|M_{\bar{p}\pi^0}-m_{\bar\Sigma^-}|<12$ MeV/$c^2$ and $|M_{\Lambda\pi^0}-m_{\Xi^0}|<12$ MeV/$c^2$ for $J/\psi\to\Xi^0\bar\Sigma^- K^+$, to enhance the signal purity. 
The signal purities increase to 93.2\%, 80.1\%, and 93.4\%, respectively. 
Figures~\ref{fig:somefig1}, \ref{fig:somefig2} and \ref{fig:somefig3} show the comparisons of the momenta and cosines of polar angles of all final-state particles between data and MC simulation, as well as the two-body invariant mass distributions. The consistencies between data and MC simulation are good, thereby ensuring the reliability of the signal MC samples.

The detection efficiencies of $J/\psi\to\Xi^0\bar\Lambda K^0_S$, $J/\psi\to\Xi^0\bar\Sigma^0 K^0_S$, and $J/\psi\to\Xi^0\bar\Sigma^- K^+$ are determined to be $(0.926\pm 0.005)\%$, $(0.144\pm 0.002)\%$, and $(0.704\pm 0.004)\%$, respectively. Efficiency correction factors have been applied to account for the data-MC deviation arising from tracking and PID efficiencies for $K$ and $p$, as well as the $\Lambda$ reconstruction, as listed in Section ~\ref{sec:sys}.

\subsection{Branching fractions}

The branching fractions of $J/\psi\to\Xi^0\bar\Lambda K^0_S$, $J/\psi\to\Xi^0\bar\Sigma^0 K^0_S$, and $J/\psi\to \Xi^0\bar\Sigma^- K^+$ are determined as:
\begin{equation}
  \mathcal{B}_{\rm{sig}} = \frac{N_{\rm{sig}}}{N_{J/\psi}\mathcal{B}_{\rm prod} \epsilon_{\rm sig}},
\end{equation}
where $N_{\rm{sig}}$ is the signal yield of each signal decay extracted from the fit to the data sample taken at $\sqrt{s}=$ 3.097 GeV. The total number of $J/\psi$ events is indicated as $N_{J/\psi}$, and $\mathcal{B}_{\rm prod}$ is the product of the branching fractions of the corresponding intermediate states~($\mathcal{B}(K_S^0\to\pi^+\pi^-)=(69.20\pm 0.05)\%$, $\mathcal{B}(\pi^0\to\gamma\gamma)=(98.823\pm 0.034)\%$, $\mathcal{B}(\bar\Lambda\to p\pi^-)=(64.1\pm 0.5)\%$, $\mathcal{B}(\bar\Sigma^0\to \gamma\bar\Lambda)=100\%$, $\mathcal{B}(\Xi^0\to \Lambda\pi^0)=(99.542\pm 0.012)\%$, $\mathcal{B}(\bar\Sigma^-\to \bar{p}\pi^0)=(51.47\pm 0.30)\%$) taken from the PDG~\cite{pdg}. The symbol $\epsilon_{\rm sig}$ stands for the detection efficiency based on MC simulation. The branching fractions of the $J/\psi\to\Xi^0\bar\Lambda K^0_S$, $J/\psi\to\Xi^0\bar\Sigma^0 K^0_S$, and $J/\psi\to\Xi^0\bar\Sigma^- K^+$ are determined to be $(3.76\pm 0.14)\times 10^{-5}$, $(2.24\pm 0.32)\times 10^{-5}$, and $(5.64\pm 0.17)\times 10^{-5}$, where the uncertainties are statistical only.

\section{SYSTEMATIC UNCERTAINTIES}\label{sec:sys}

The systematic uncertainties in the measurements of the branching fractions of $J/\psi\to\Xi^0\bar\Lambda K^0_S$, $J/\psi\to\Xi^0\bar\Sigma^0 K^0_S$, and $J/\psi\to\Xi^0\bar\Sigma^- K^+$ are discussed below.

The total number of $J/\psi$ events in data has been determined to be $N_{J/\psi}=(10087\pm 44)\times 10^6$ with inclusive hadronic events as described in Ref.~\cite{jpsinumber}. This entails a systematic uncertainty of 0.4\%.

The systematic uncertainties of the $K$ tracking and PID are studied with the control sample of $\psi(3686)\to \pi^+\pi^-J/\psi$ with $J/\psi\to 2(K^+K^-)$~\cite{syskaon}. The differences of $K$ tracking and PID efficiencies between data and MC simulation are obtained in different transverse momentum intervals. The data-MC differences are then weighted according to the distribution of the transverse momentum of kaon in the signal decay. The data to MC ratios of the re-weighted tracking and PID efficiencies are $(99.7\pm 0.4)\%$ and $(99.5\pm 0.1)\%$, respectively. Here the errors originate mainly from the limited statistics of the control sample. The detection efficiency estimated from the signal MC sample is corrected with the data to MC ratios, and the residual uncertainties of the ratios, $0.4\%$ and $0.1\%$, are taken as the systematic uncertainties of the tracking and PID per $K$, respectively.

The systematic uncertainties of $p$ tracking and PID are studied with the control sample of $J/\psi\to p\bar{p}\pi^+\pi^-$. The differences of $p$ tracking and PID efficiencies between data and MC simulation are obtained in different two dimensional intervals of polar angle and transverse momentum. The data to MC ratios of the re-weighted tracking and PID efficiencies are $(100.0\pm 0.2)\%$ and $(100.7\pm 0.2)\%$, respectively. After weighting them according to the distribution of the polar angle and transverse momentum of $p$ in the signal decay, we correct the MC efficiency and the residual uncertainties, $0.2\%$ is taken as the systematic uncertainty of both tracking and PID per $p$.

The photon efficiency is studied with the sample of $\psi(3686)\to\pi^+\pi^-J/\psi$, $J/\psi\to\rho^0\pi^0$ by the  three methods listed in Ref.~\cite{sysphoton}. The difference of photon detection efficiency between data and MC simulation, 1.0\%, is taken as the systematic uncertainty.

The efficiencies of $\Lambda$ reconstruction, including the tracking efficiencies of the $p \pi^-$ pair, decay length cut, mass window cut, vertex fit and secondary vertex fit, are studied by using the control samples of $J/\psi\to pK^-\bar{\Lambda}$ and $J/\psi\to \Lambda\bar{\Lambda}$~\cite{syslambda}. The obtained efficiencies in the 2D intervals of momentum and $\cos\theta$ are then weighted to match our signal MC events. The differences of the re-weighted efficiencies between data and MC simulation are $(96.6\pm 1.0)\%$ for $\Lambda$ and $(95.8\pm 1.0)\%$ for $\bar\Lambda$ in $J/\psi\to\Xi^0\bar\Lambda K^0_S$, $(93.8\pm 1.2)\%$ for $\Lambda$ and $(91.9\pm 1.3)\%$ for $\bar\Lambda$ in $J/\psi\to\Xi^0\bar\Sigma^0 K^0_S$, and $(91.3\pm 1.0)\%$ in $J/\psi\to\Xi^0\bar\Sigma^- K^+$, respectively. We correct the $\Lambda$ reconstruction efficiency of MC simulation to data and assign $2.0\%$, $2.5\%$ and $1.0\%$ as the uncertainties of the correction factors.

The efficiency of the $K_S^0$ reconstruction incorporating the tracking and the mass window selection is studied using the control sample of $J/\psi\to K^*(892)^\pm K^\mp$, with $K^*(892)^\pm\to K_S^0\pi^\pm$. The difference of $K_S^0$ reconstruction efficiency between data and MC simulation, 2\%, is taken as the systematic uncertainty.

The systematic uncertainty due to the $\pi^0$ selection is determined from a high purity control sample of $J/\psi\to \pi^+\pi^-\pi^0$. The difference of $\pi^0$ reconstruction efficiency between data and MC simulation gives an uncertainty of $1\%$ per $\pi^0$~\cite{pi0sys}, which includes the $\gamma$ reconstruction.

The systematic uncertainty of the 2D fit is considered in two aspects.
The background shape is changed from a second to either a first or a third order polynomial.
The signal shape is changed to remove the convolved Gaussian function. According to Ref.~\cite{rbmethod}, taking the correlations between samples into account, any deviation within $\pm2\sigma$ of the nominal yield is deemed negligible. Applying this criterion, we assign a 2.9\% systematic uncertainty to the $J/\psi\to\Xi^0\bar\Lambda K^0_S$ channel, while those for the other two channels are negligible.

In the channel $J/\psi\to\Xi^0\bar\Lambda K^0_S$, the decay $J/\psi\to\Xi^0\bar\Sigma^0 K^0_S$ produces a fully peaking background in the mass distributions. To estimate the effect of this background, we calculate the misidentification rate using the signal MC sample of $J/\psi\to\Xi^0\bar\Sigma^0 K^0_S$ that passes the event selection criteria for $J/\psi\to\Xi^0\bar\Lambda K^0_S$. The misidentification rate is 0.001\%. Combining this with the signal efficiency of $J/\psi\to\Xi^0\bar\Lambda K^0_S$ and the branching fraction obtained in this analysis, we assign a conservative uncertainty of 2.0\% to account for the influence of this background.

In the nominal analysis, the helix parameters of charged tracks in the 5C or 6C kinematic fit have been corrected with the parameters derived with the control sample of $J/\psi\to \phi f_0(980), \phi\to K^+K^-, f_0(980)\to \pi^+\pi^-$~\cite{helixkpi}, and $\psi(3686)\to \pi^0 p\bar{p}$. The differences of detection efficiencies with and without helix parameter correction, $2.6\%$, $8.8\%$, and $2.0\%$ for the three signal channels, are assigned as the corresponding systematic uncertainties. The uncertainty for the second channel is larger than the other two mainly due to the miscombinations of signal candidate.

For each signal decay, the uncertainty due to the MC statistics is defined by
\begin{equation}\centering
  \frac{1}{\sqrt{N}}\sqrt{\frac{(1-\epsilon)}{\epsilon}},
\end{equation}
where $\epsilon$ is the detection efficiency and $N$ is the total number of signal MC events. The assigned systematic uncertainty is 0.1\%.

The $\Lambda$, $\Sigma^+$, $\Xi^0$ are generated in the PHSP in the nominal analysis. Alternative signal MC samples are generated with the decay parameters quoted from PDG~\cite{pdg} containing $S\text{-wave}$ and $P\text{-wave}$. The relative changes in detection efficiencies, 1.6\%, 0.1\%, 1.1\%, are assigned as the systematic uncertainties for MC model.

The branching fractions ${\mathcal B}_{K_S^0\to \pi^+\pi^-}=(69.2\pm 0.05)\%$, ${\mathcal B}_{\Lambda\to p\pi^-}=(64.1\pm 0.5)\%$, and ${\mathcal B}_{\bar\Sigma^-\to \bar{p}\pi^0}=(51.5\pm 0.3)\%$ are quoted from the PDG~\cite{pdg}. They contribute with uncertainties of 1.6\%, 1.6\%, and 1.4\% for $J/\psi\to \Xi^0\bar\Lambda K^0_S$, $J/\psi\to \Xi^0\bar\Sigma^0 K^0_S$, and $J/\psi\to \Xi^0\bar\Sigma^- K^+$, respectively.

All systematic uncertainties are summarized in Table~\ref{tab:Sys}. Assuming that all sources are independent, the total systematic uncertainties for each signal decay are calculated by adding all systematic uncertainties quadratically.

\begin{table}[htbp]\centering
  \caption{Systematic uncertainties (in unit of \%) in the branching fraction measurements. Symbol denoted by ``..." indicates that no systematic uncertainty was considered.}
  \begin{tabular}{lccc}
      \hline\hline
      Source                    & $\Xi^0\bar\Lambda K^0_S$ & $\Xi^0\bar\Sigma^0 K^0_S$ & $\Xi^0\bar\Sigma^- K^+$ \\
      \hline
      $N_{J/\psi}$              & 0.4  & 0.4   & 0.4  \\
      Tracking                  & ...  & ...   & 0.6  \\
      PID                       & ...  & ...   & 0.3  \\
      $\gamma$ detection        & ...  & 1.0   & ...  \\
      $K_S^0$ reconstruction    & 2.0  & 2.0   & ...  \\
      $\Lambda$ reconstruction  & 2.0  & 2.5   & 1.0  \\
      $\pi^0$ reconstruction    & 2.0  & 2.0   & 4.0  \\
      2D fit                    & 2.9  & ...   & ...  \\
      Kinematic fit             & 2.6  & 8.8   & 2.0  \\
      MC statistics             & 0.1  & 0.1   & 0.1  \\
      MC model                  & 1.6  & 0.1   & 1.1  \\
      Quoted branching fractions & 1.6  & 1.6   & 1.4  \\
      Peaking background        & 2.0  & ...   & ...  \\
      \hline
      Total                     & 6.0  & 9.8   & 5.0 \\
      \hline\hline
  \end{tabular}
  \label{tab:Sys}
\end{table}

\section{SUMMARY}\label{sec:summary}

By using $(10087 \pm 44)\times10^6$ $J/\psi$ events taken at the BESIII detector, we have measured the branching fractions of the decays $J/\psi\to\Xi^0\bar\Lambda K^0_S$, $J/\psi\to\Xi^0\bar\Sigma^0 K^0_S$, and $J/\psi\to \Xi^0\bar\Sigma^- K^+$ for the first time to be:
\begin{align*}
 \mathcal{B}(J/\psi\to\Xi^0\bar\Lambda K^0_S) &= (3.76\pm0.14\pm 0.22)\times10^{-5}, \\
 \mathcal{B}(J/\psi\to\Xi^0\bar\Sigma^0 K^0_S) &= (2.24\pm0.32\pm 0.22)\times10^{-5}, \\
 \mathcal{B}(J/\psi\to\Xi^0\bar\Sigma^- K^+) &= (5.64\pm0.17\pm 0.27)\times10^{-5},
\end{align*}
where the first uncertainties are statistical and the second systematic. The ratio of the branching fractions between $J/\psi\to\Xi^0\bar\Sigma^0 K^0_S$ and $J/\psi\to\Xi^0\bar\Sigma^- K^+$ is determined to be $\mathcal{R}=0.40 \pm 0.07$, which deviates by $2.1\sigma$ from the isospin conservation number $1/4$. Here, the correlated systematic uncertainties associated with the reconstruction of $\Lambda$ and $\pi^0$, kinematic fit, MC statistics, as well as the quoted BF and $N_{J/\psi}$, largely cancel out. When combined with the branching fractions of the $\psi(3686)$ decays into the same final states, which will be obtained at BESIII in the near future, they will offer valuable information to understand the $\rho\pi$ puzzle in the $\psi$ decays. The excited hyperon, $\Xi(1690)^{0}$, can be seen in the $M_{K^{-} \Sigma^+}$ distributions, where a potential $\Xi(1720)^{0}$ contribution is also indicated. However, due to the limited statistics and the large systematic uncertainty arising from the low momentum $\Lambda$, no definitive conclusion can be obtained. No baryon-anti-baryon threshold enhancement is observed.

\textbf{Acknowledgement}

The BESIII Collaboration thanks the staff of BEPCII (https://cstr.cn/31109.02.BEPC) and the IHEP computing center for their strong support. This work is supported in part by National Key R\&D Program of China under Contracts Nos. 2023YFA1606000, 2023YFA1606704; National Natural Science Foundation of China (NSFC) under Contracts Nos. 11635010, 11935015, 11935016, 11935018, 12025502, 12035009, 12035013, 12061131003, 12192260, 12192261, 12192262, 12192263, 12192264, 12192265, 12221005, 12225509, 12235017, 12361141819; the Chinese Academy of Sciences (CAS) Large-Scale Scientific Facility Program; the Strategic Priority Research Program of Chinese Academy of Sciences under Contract No. XDA0480600; CAS under Contract No. YSBR-101; 100 Talents Program of CAS; The Institute of Nuclear and Particle Physics (INPAC) and Shanghai Key Laboratory for Particle Physics and Cosmology; ERC under Contract No. 758462; German Research Foundation DFG under Contract No. FOR5327; Istituto Nazionale di Fisica Nucleare, Italy; Knut and Alice Wallenberg Foundation under Contracts Nos. 2021.0174, 2021.0299; Ministry of Development of Turkey under Contract No. DPT2006K-120470; National Research Foundation of Korea under Contract No. NRF-2022R1A2C1092335; National Science and Technology fund of Mongolia; Polish National Science Centre under Contract No. 2024/53/B/ST2/00975; STFC (United Kingdom); Swedish Research Council under Contract No. 2019.04595; U. S. Department of Energy under Contract No. DE-FG02-05ER41374


\end{document}